\input mtexsis.tex
\input epsf.tex
\paper
\tenpoint


\superrefsfalse
\referencelist
\reference{Bardeen} {Bardeen, J.} {\sl"A variational principle for
rotating stars in general relativity."} {\journal ApJ;162,
71-95(1970)}\endreference

\reference{Bicak} {B\`\i c\'ak, J. \& Ledvinka, T. } {\sl"Relativistic disks
as sources of the Kerr metric."} {\journal Phys Rev D;71,
1669-1672(1993)}\endreference

\reference{BLK} {B\`\i c\'ak, J. Lynden-Bell, D. et al.}
{\sl"Relativistic disks as sources of static vacuum spacetimes."}
{\journal Phys Rev D;47, 4334-4343(1993)}\endreference

 \reference{BLP} {B\`\i c\'ak J, Lynden-Bell, D \& Pichon
C.}{\sl``Relativistic discs and flat galaxy models''}{\journal
MNRAS;265,126-144(1993)}\endreference 

\reference{Chamorow} {Chamorow, A., R. Gregory, et al.} {\sl"Static
axisymmetric discs and gravitational collapse."} {\journal Proc
R. Soc. Lond.;413, 251-262(1987)}\endreference

\reference{Curzon} {Curzon,H.} \journal Proc.Lond.Math.Soc.;23,477-480(1924)\endreference

\reference{Dietz} {Dietz, W. \& C. Hoenselaers} {\sl"Stationary
system of two masses kept apart by their gravitational spin spin
interaction."} {\journal Phys Rev Lett;48, 778-780(1982)}\endreference

\reference{Dietz2} {Dietz, W. and C. Hoenselaers} {\sl"Two mass
Solutions of Einstein's Vacuum Equations: The double Kerr solution."}
{\journal Ann Phys;165, 319-383(1985)}\endreference

\reference{Evans} {Evans, N. and P. de Zeeuw} {\sl"Potential-density
Pairs for Flat Galaxies."} {\journal MNRAS;257,
152-176(1992)}\endreference

\reference{Fackerell} {Fackerell, E.} {} {\journal ApJ;153,
643-660(1968)}\endreference

\reference{Katz} {Katz, J.  Horwitz, G \& Klapish, M.} 
\journal ApJ;199,307-321(1975)\endreference

\reference{Kerr} {Kerr,R.} \journal Phys.Rev.Let.;11,237-238(1963)\endreference
\reference{HKX} {Hoenselaers, C., W. Kinnersley, et al.}
{\sl"Symmetries of the stationary Einstein-Maxwell equations, VI
Transformations which generate asymptotically flat spacetimes with
arbitrary multipole moments."} {\journal J~Math~Phys;20,
2530-2536(1979)}\endreference

\reference{HKX2} {Hoenselaers, C. and W. Dietz} {\sl"The rank N HKX
Transformations: New stationary Axisymmetric Gravitational fields."}
{\journal Gen Rel Grav;16, 71-78(1984)}\endreference

\reference{Ipser} {Ipser,J.} {}
 {\journal ApJ;158, 17-43(1969)}\endreference

\reference{Israel}{ Israel, W.} \journal Nuovo-Cimento;33,331-335(1964)\endreference
\reference{Lemos} {Lemos, J.} {\sl"Self Similar relativistic Discs
with Pressure."} {\journal Class Q Grav;6,
1219-1230(1989)}\endreference

\reference{Morgan}{ Morgan,T, Morgan L.} \journal Phys.Rev.;183,1097-1102(1969)\endreference

\reference{Gravitation}{Misner, C.,Thorne, K. \& Wheeler, J. }, {\sl 
Gravitation},Freeman,(1973)\endreference

\reference{PDLB}{Pichon, C, \& Lynden-Bell, D}{\sl"Equilibria of flat
and round disks"}(1995){ to appear in MNRAS} \endreference

\reference{Synge}{ Synge,J.} {\sl The Relativistic Gas}, North Holand (1957) \endreference

\reference{Thorne} {Ipser, J. \& Thorne} {}
 {\journal ApJ;154, 251-(1968)}\endreference

\reference{Tolman} {Tolman, R.} {\sl"Relativity, Thermodynamics and
Cosmology."}, Oxford University Press, {(1934)}\endreference

\reference{Toomre} {Toomre, A.} {\sl"On the Gravitational Stability of
a Disc of Stars."} {\journal ApJ;139, 1217-1238(1964)}\endreference

\reference{Voorhess} {Voorhess, B.} \journal Phys. Rev. {\bf D}.;2,2119(1970)\endreference

\reference{Weyl}{Weyl, H.} \journal Ann.Phys.;54,117-118(1917)\endreference

\reference{Yamazaki} {Yamazaki, M.} {\sl"On the
Hoenselaers-Kinnersley-Xanthopoulos spinning mass fields.} {\journal
  J Math. Phys.;22, 133-135(1981)}\endreference

\endreferencelist


\def\Note{\relax}

\titlepage 
\title{New sources for Kerr and other metrics:
rotating relativistic disks with pressure support.}
\author
C. Pichon$^{1,2}$, \& D. Lynden-Bell$^1$ 
\centerline{$1$ Institute of Astronomy, Madingley Road, Cambridge CB3 0HA}
\centerline{$2$ CITA,McLennan Labs,University of Toronto,}
\centerline{60 St. George Street,Toronto,Ontario M5S 1A7.}
\endauthor

\abstract

 Complete sequences of new analytic solutions of Einstein's equations
which describe thin super massive disks are constructed.  These
solutions are derived geometrically.  The identification of points
across two symmetrical cuts through a vacuum solution of Einstein's
equations defines the gradient discontinuity from which the properties
of the disk can be deduced.  The subset of possible cuts which lead to
physical solutions is presented.  At large distances, all these disks
become Newtonian, but in their central regions they exhibit
relativistic features such as velocities close that of light, and large
redshifts.  
Sections with zero extrinsic curvature yield cold disks.  Curved
sections may induce disks which are stable against radial instability.
The general counter rotating flat disk with planar pressure tensor is
found.  Owing to gravomagnetic forces, there is no  systematic method
of constructing vacuum stationary fields for which the non-diagonal
component of the metric is a free parameter.  However, all static vacuum
solutions may be extended to fully stationary fields via simple
algebraic transformations. Such disks can generate a great variety of
different metrics including Kerr's metric with any ratio of $a$ to
$m$.  A simple inversion formula is given  which yields all
distribution functions compatible with the characteristics of the flow,
providing formally  a complete description of the stellar dynamics of
flattened relativistic disks. It is illustrated for the Kerr disk.
\vskip 1cm

\raggedcenter{\ninepoint {\bf keywords}: Relativity -- methods: analytical
--  stellar dynamics, disks  --  quasars}
\endraggedcenter

\centerline{\sl accepted by the 
Monthly Notices of the Royal Astronomical Society,
 (1996){ \bf 280 (4)},1007-1026.
}
\endtitlepage

\section{Introduction}

The general-relativistic theory of rapidly rotating objects is of
 great intrinsic interest and has potentially important applications
 in astrophysics.  Following Einstein's early work on relativistic
 collisionless spherical shells, Fackerell (1968)\cite{Fackerell}, Ipser
 (1969)\cite{Ipser} and Ipser \& Thorne (1968)\cite{Thorne} have
 developed the general theory of relativistic collisionless spherical
 equilibria.  The stability of such bodies has been further expanded
 by Katz (1975)\cite{Katz}. Here the complete dynamics of
 flat disk equilibria is developed and illustrated, extending these
 results to differentially rotating configurations with partial
 pressure support. A general inversion method for the corresponding
 distribution functions is presented, yielding a coherent model for
 the stellar dynamics of these disks.  Differentially rotating flat
 disks in which centrifugal force almost balances gravity can also
 give rise to relatively long lived configurations with large binding
 energies.  Such objects may therefore correspond to possible models
 for the latest stage of the collapse of a proto-quasar.  Analytic
 calculations of their structure have been carried into the post
 Newtonian r\'egime by Chandrasekhar. Strongly relativistic bodies
 have been studied numerically following the pioneering work on
 uniformly rotating cold disks by Bardeen \& Wagoner\cite{Bardeen}.
 An analytic vacuum solution to the Einstein equations, the Kerr
 metric\cite{Kerr}, which is asymptotically flat and has the general
 properties expected of an exterior metric of a rotating object, has
 been known for thirty years, but attempts to fit an interior solution
 to its exterior metric have been unsatisfactory.  A method for
 generating families of self-gravitating rapidly rotating disks purely
 by geometrical methods is presented. Known vacuum solutions of
 Einstein's equations are used to produce the corresponding
 relativistic disks.  One family includes interior solutions for the
 Kerr metric.

\subsection{ Background}

In 1919 Weyl\cite{Weyl} and Levi-Civita gave a method for finding all
solutions of the axially symmetric Einstein field equations after
imposing the simplifying conditions that they describe a static vacuum
field.  B\`\i c\'ak, Lynden-Bell and Kalnajs (1993)\cite{BLK} -- here-after
refered as BLK -- derived complete solutions corresponding to counter
rotating pressure-less axisymmetric disks by matching exterior
solutions of the Weyl- Levi-Civita type on each side of the disk.  For
static metrics, the vacuum solution was obtained via line
superposition of fictitious sources placed symmetrically on each side
of the disk by analogy with the classical method of images.  The
corresponding solutions describe two counter-rotating stellar streams.
This method will here be generalised, allowing {\sl true rotation} and
radial pressure support for solutions describing self-gravitating
disks.  These solutions provide physical sources (interior solutions)
for stationary axisymmetric gravitational fields.

\subsubsection{ Pressure Support via Curvature }

In Newtonian theory, once a solution for the vacuum field of Laplace's
equation is known, it is straightforward to build solutions to
Poisson's equation for disks by using the method of mirror images.  In
general relativity, the overall picture is similar.  In fact, it was
pointed out by Morgan \& Morgan (1969)\cite{Morgan} that for
pressureless counter rotating disks in Weyl's co-ordinates, the
Einstein equation reduces essentially to solving Laplace's equation.
  When pressure is present, no global set of
Weyl co-ordinates exists.  The condition for the existence of such a
set of co-ordinates is that $T_{R R}+T_{z z}=0$.  On the disk itself,
this is satisfied only if the radial eigenvalue of the pressure tensor,
$p_R$, vanishes ($T_{z z}$ vanishes provided the disk is thin).  When
$p_R$ is non-zero, vacuum co-ordinates of Weyl's type still apply both
above and below the disk, but as two separate co-ordinate patches.
Consider global axisymmetric co-ordinates $(R',\phi',z')$ with $z'=0$
on the disk itself. For $z ' >0$, Weyl's co-ordinates $R,\phi,z$ may
be used.  In terms of these variables, the upper boundary of the disk
$z'=0^+$ is mapped onto a given surface $z=f(R)$. On the lower patch,
by symmetry, the disk will be located on the surface $z=-f(R)$. Thus
in Weyl co-ordinates, the points $z=f(R)$ on the upper surface of the
disk must be identified with the points $ z=-f(R)$ on the lower
surface.  The intrinsic curvature of the two surfaces that are
identified match by symmetry.  The jump in the extrinsic curvature
gives the surface distribution of stress-energy on the disk. A given
metric of Weyl's form is a solution of the empty space Einstein
equations, and does not give a complete specification of its sources
(for example, a Schwarzschild metric of any given mass can be
generated by a static spherical shell of any radius).  With this
method, all physical properties of the source are entirely
characterised once it is specified that the source lie on the surface
$z= \pm f(R)$ and the corresponding extrinsic curvature (i.e. its
relative layout within the embedding vacuum spacetime) is known.  The
freedom of choice of $\nu$ in the Weyl metric corresponds to the
freedom to choose the density of the disk as a function of radius. The
supplementary degree of freedom involved in choosing the surface of
section $z=f(R)$ corresponds classically to the choice of the radial
pressure profile.
\subsubsection{ Rotating Disks}
 
For rotating disks, the dragging of inertial frames induces strongly
non-linear fields outside the disk which prohibit the construction of
a vacuum solution by superposition of the line densities of fictitious
sources.  nevertheless, in practise quite a few vacuum stationary
solutions have been given in the literature, the most famous being the
Kerr metric.  Hoenselaers, Kinnersley \& Xanthopoulos
(1979)\cite{HKX}\cite{HKX2} (HKX hereafter) have also given a discrete
method of generating rotating solutions from known static Weyl
solutions.  The disks are derived by taking a cut through such a field
above all singularities or sources and identifying it with a
symmetrical cut through a reflection of the source field.  This method
applies directly to the non-linear fields such as those generated by a
rotating metric of the Weyl-Papapetrou type describing stationary
axisymmetric vacuum solutions.  The analogy with electromagnetism is
then to consider the field associated with a known azimuthal vector
potential $A_\phi$, the analog of $g_{t \phi}/g_{t t}$. The jump in the
tangential component of the corresponding $B$-field gives the electric
current in the plane, just as the jump in the $t,\phi$ component of
the extrinsic curvature gives the matter current.
 
The procedure is then the following: for a given surface embedded in
curved space-time and a given field outside the disk, the jump in the
extrinsic curvature on each side of the mirror images of that surface
is determined in order to specify the matter distribution within the
disk.  The formal relationship between extrinsic curvatures and
surface energy tensor per unit area was introduced by Israel and is
described in details by Misner Thorne Wheeler (1973) p
552.\cite{Gravitation} Plane surfaces and non-rotating vacuum fields
(\ie the direct counterpart of the classical case) were investigated
by BLK  and led to pressureless counter-rotating
disks.  Any curved surface will therefore produce disks with some
radial pressure support.  Any vacuum metric with $g_{t \phi} \not= 0$
outside the disk will induce rotating disks.

\midfigure{cut}
\Caption
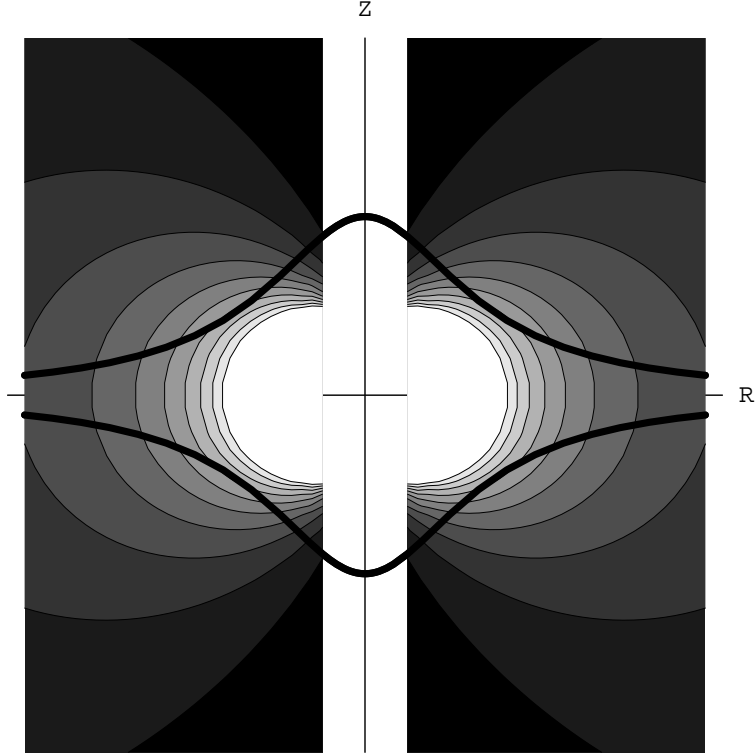
Symbolic representation of a section $z=\pm f(R)$ through the $g_{t 
\phi}$
field  representing the upper 
and lower surface of the disk embedded in two Weyl-Papapetrou fields.
The isocontours of $\omega(R,z)$  
correspond to a measure of the amount of rotation in the disk. 
\endCaption
\endfigure 
  
 In section 2, the construction scheme for rotating disks with
pressure support is presented.  In section 3, the properties of these
disks are analysed, and a reasonable cut, $z=f(R)$, is suggested.
Section~4 describes stellar equilibria which have 
the corresponding anisotropic stress energy tensors, while section~5 
derives the  properties of the  limiting counter-rotating disks.
This method is then first applied to warm counter rotating disks in
section 6; the dominant energy condition given for Curzon disks by
Chamorow {\it et. al.}(1987)\cite{Chamorow}, and Lemos ' solution for
the self - similar Mestel disk with pressure support are recovered.
In section 7, the Kerr disk is studied and the method for producing
general HKX disks is sketched.

\section{Derivation}

The jump in extrinsic curvature of a given profile is calculated and
related to the stress energy tensor of the matter distribution in the
corresponding self gravitating relativistic disk.
\subsection{ The Extrinsic Curvature }

The line element corresponding to an axisymmetric 
stationary vacuum 
  gravitational field 
is given by  the Weyl-Papapetrou  (WP) metric 
$$ \eqalign{ ds^2 = &
  - {e^{2\,\nu }}\,{{{\left(\it dt -\omega \, d\phi\right)}}^2}  
 +{e^{ 2\,\zeta-2\,\nu   }}\,( {{{\it
  dR}}^2} +{{{\it dz}}^2}) + {{}{R^2}{{e^{-2\,\nu }}}\, d \phi^2},  \cr}
\EQN WeylPmetric
  $$

where $(R,\phi,z)$ are standard cylindrical co-ordinates, and $\nu$, $\zeta$ and 
$\omega$
are functions of $(R,z)$ only\Note{geometric units $G=c=1$ are used throughout this paper}.
 Note that this form is explicitly symmetric under 
simultaneous 
change of $\phi$ and $t$. 
The vacuum field induces a natural metric on a 3-space-like $z =f(R)$ 
surface
$$\eqalign{ d\,\sigma ^2
=g_{\alpha \beta }\,{{\partial x^\alpha }
 \over {\partial x^a}}{{\partial x^\beta }
 \over {\partial x^b}}dx^a dx^b &\equiv g_{\alpha \beta }
\,h_a^\alpha h_b^\beta \,dx^adx^b, \cr &\equiv \gamma _{ab}\,dx^a\,dx^b ,
\cr } \EQN diskmetric $$
where $\{ x^a \}$ are  the co-ordinates on the embedded hypersurface
 $z= f(R)$,  $\{ x^a \}= ( t,R,\phi)$, and  $\{ x^\alpha  \}=( t,R,z,\phi)$ are Weyl's
co-ordinates for the embedding spacetime. Here, $\gamma_{a b}$ stands for 
the embedded metric, and $h_a^\alpha$ is given by
$$h_a^\alpha =
{{\partial x^\alpha } \over {\partial x^a}}
=\left[ \matrix{ 1 & 0 & 0 & 0 \cr 0 & 1 & 0 & f' \cr
  0 & 0 & 1 & 0 \cr  } \right], \EQN h $$
where hereafter a prime, $(\,\,)'$, represents  the derivative with respect to $R$. 
The line element is then 
$$\eqalign{ d \sigma^2=& - {e^{2\,\nu }}\,{{{\left(\it dt -\omega \, d\phi\right)}}^2} 
+ {{{R^2}} {{e^{-2\,\nu }}}}{{{
  d\phi}}^2} + {e^{ 2\,\zeta-2\,\nu  }}\,\left( 1 + f'^2 \right) 
 \, {{{\it dR}}^2}. }\EQN diskmetric $$

Let  the upward pointing normal $N_\mu $ to the surface  $z= f(R)$ be
$$N_\mu ={{\partial } \over {\partial x^\mu }}\left[z-f(R)\right]=(0,-f',0,1).
\EQN normal$$
The normalised normal is therefore
$$n_\mu ={{N_\mu } \over {\sqrt {N_\alpha N^\alpha }}}=
{{N_\mu } \over {\sqrt {g^{\alpha \beta }N_\beta N_\alpha }}}=
{{N_\mu } \over {\sqrt {1+f'^2}}}e^{\zeta-\nu } .
\EQN Nnormal$$
The extrinsic curvature tensor $K$ is given in its covariant form by
$$  K_{a b}=n_\mu \left( {\partial _ah_b^\mu +
\Gamma _{\alpha \,\beta }^\mu \,h_a^\alpha \,h_b^\beta } \right),
\EQN K$$
 where 
$\partial _ah_b^\mu =
{{\partial h_b^\mu } / {\partial x^a}}.$
\Eq{Nnormal} and  \Ep{h} together with the Christoffel symbols for the
Weyl metric  lead to 
the computation of $K$ ,
  the extrinsic curvature of that surface embedded in the Weyl Papapetrou metric:
$$\EQNalign{
&K_{tt}=-{{e^{\nu -\zeta }} \over {\sqrt {1+f'^2}}}
\left( {{{\partial \nu } \over {\partial N}}} \right) e^{ 4 \nu -2 \zeta} ,\EQN Kdown;a \cr
  &K_{RR}=-{{e^{\nu -\zeta }} \over { \sqrt{1+f'^2}}}
\left( {{{f''} \over { {1+f'^2}}}+
{{\partial (\nu -\zeta )} \over {\partial N}}} \right)(1+f'^2),\EQN Kdown;b \cr
  &K_{\phi t} =-{{e^{\nu -\zeta }} \over {\sqrt {1+f'^2}}}
\left[ {{\partial [\nu +\half\log(\omega)] } \over {\partial N}} \right]
 \omega e^{ 4 \nu -2 \zeta} ,\EQN Kdown;c \cr
&
K _{\phi\phi} =-{{e^{\nu -\zeta }} \over {\sqrt {1+f'^2}}}
\left[
\left( {{{f'} \over R}+{{\partial \nu } \over {\partial N}}} \right) R^2 e^{-2\zeta}
 +e^{4 \nu -2 \zeta} \omega^2 {{\partial [\nu -\log(\omega) ]} \over {\partial 
N}}
\right], \EQN Kdown;d
\cr  }
 $$
where the notation 
${\partial  / {\partial N}} 
\equiv
\left( {{\partial  / {\partial z}}} -f'
 {{\partial  / {\partial R}}} \right) $ has been used.


\subsection{ The Stress Energy Tensor}

The stress energy tensor per unit  surface $\tau_b^a$ is 
the integral of of the stress energy tensor carried along
the normal to the surface $z=f ( R )$.
In a locally Minkowskian frame co-moving with the mean flow of the 
disk, the corresponding orthonormal tetrad is 
$$\EQNalign{&e_i^{(0)}=e^\nu (1,0,-\omega ), \EQN tetrad;a  \cr 
  &e_i^{(1)}=e^{\zeta -\nu }\sqrt {1+f'^2}(0,1,0), \EQN tetrad;b \cr
  &e_i^{(2)}= R e^{-\nu} (0,0,1), \EQN tetrad;c \cr} $$
so that 
$$ d \,s^2= \eta_{(a)(b)} \left(e_i^{(a)} d\, x^i \right)  \left(e_j^{(b)} d\, x^j \right)
,\EQN metric tetrad$$
with $ \eta_{(a)(b)} ={\rm Diag}(-1,1,1)$.
In that frame,
$$\left[ {\tau ^{(a)(b)}} \right]_0=\left[ {\matrix{\varepsilon &0&0\cr
0&{\,p_{_R}}&0\cr
0&0&{p_\phi }\cr
}} \right]. \EQN tau0$$
After a Lorentz transformation to a more general frame in which the flow 
is  rotating with relative velocity $V$ in the $\phi$ direction,  this stress becomes
$$\eqalign{ &
 {\tau ^{(a)(b)}} =  
{1 \over {1-V ^2}}\left[ {\matrix{{\varepsilon +p_\phi \,V 
^2}&
0&{(p_\phi +\varepsilon )\,V }\cr
0&{(1-V ^2)\,p_R}&0\cr
{(p_\phi +\varepsilon )\,V }&0&{{p_\phi +\varepsilon }V ^2}\cr
}} \right] \cr} . \EQN Tau$$

\subsection{ The Discontinuity Equations}

Israel (1964)\cite{Israel} has shown that Einstein's equations  integrated through a given 
surface of discontinuity can be re-written in terms of the jump in extrinsic curvature, 
namely 
$$\ \tau_b^a= {1\over 8 \pi} \left[ {K_b^a -K^a_{a\,}\,\delta_b^a} \right]^+_- 
\equiv  {\cal L}_b^a .\EQN T2K$$
where $ [\quad]^+_-$ stands for $(\quad)$ taken on $z=f(R)$ minus 
$(\quad)$ taken on  $z=-f(R)$. 
This tensor ${\cal L}_b^a $ is known as the Lanczos tensor.

In the tetrad frame \Ep{ tetrad},  the Lanczos tensor given by \Eqs{ T2K}
 and \Ep{ Kdown } reads
 $$ \eqalign{ &
{\cal L}^{(a)(b)} =  {{e^{\nu -\zeta }} \over {4\pi \sqrt {1+f'^2}}}  
 \left[ {\matrix{{{{f''} \over {1+f'^2}}+{{f'} \over R}+{{\partial (2\nu -\zeta )} 
\over {\partial N}}}&0&{{{\partial \omega } \over {\partial N}}{{e^{2\nu }} \over 
{2R}}}\cr
0&{-{{f'} \over R}}&0\cr
{{{\partial \omega } \over {\partial N}}{{e^{2\nu }} \over {2R}}}&0&{-{{f''} \over 
{1+f'^2}}+{{\partial \zeta } \over {\partial N}}}\cr
}} \right] \cr}.
 \EQN Lanczos$$
Identifying the  stress energy tensor  $\tau^{(a)(b)}$ with the tetrad Lanczos tensor
${\cal L}^{(a)(b)}$ according to \Eq{ T2K}, and solving for
 $p_{{}_R} \, $,$\, p_\phi \, $,$\,\varepsilon\,$
 and $V$ yields
$$\EQNalign{&V ={{R\,e^{-2\nu }} \over {\partial \omega /\partial N}}
\left[ {\left( {{{f'} \over R}+2{{\partial \nu } \over {\partial N}}} \right)} -Q \right], 
\EQN FinalV;a\cr
  &\varepsilon ={{e^{\nu -\zeta }} \over {4\pi \sqrt {1+f'^2}}}\,\left[ {{Q \over 
2}+\left( {{{f''} \over {1+f'^2}}+{{f'} \over {2R}}+{{\partial (\nu -\zeta )} \over 
{\partial N}}} \right)} \right], \EQN FinalV;b\cr
  &p_\phi ={{e^{\nu -\zeta }}\over {4\pi \sqrt {1+f'^2}}}\left[ {{Q \over 2}-\left( 
{{{f''} \over {1+f'^2}}+{{f'} \over {2R}}+{{\partial (\nu -\zeta )} \over {\partial 
N}}} \right)}
 \right], \EQN FinalV;c\cr
&p_{ {}_R}={{e^{\nu -\zeta }} \over  {4\pi \sqrt {1+f'^2}}}\,\left( {{{-f'} \over R}} 
\right), \EQN FinalV;d
  \cr }  $$
where 
$$Q \equiv \sqrt {\left( {{{f'} \over R}+2{{\partial \nu } \over {\partial N}}} 
\right)^2-{{e^{4\nu }} 
\over {R^2}}\left( {{{\partial \omega } \over {\partial N}}} \right)^2}. \EQN Q$$
All quantities are to be evaluated along $z=f(R)$.
\Eq{FinalV} gives the form of  the most general solution to  the relativistic 
 rotating thin disk problem provided the expressions for $\varepsilon,p_\phi,p_{ {}_R}$ 
are physically
acceptable.

\section{ Physical properties of the warm disks}

 Physical properties of interest for these 
disks are derived  and related  to the choice of profile $z=f(R)$ compatible with 
their dynamical stability.

Defining the circumferential radius $R_c$, proper radial length  $\tilde R$, 
and a 
synchronised proper time  $\tau_*$ by  
$$\EQNalign{&R_c=R\,e^{-\nu }\,, \EQN Landau;a\cr
  &\tilde R=\int {\sqrt {1+f'^2}e^{\zeta -\nu }}\,dR \, ,\EQN Landau;b\cr
  &\tau _*=\int_{{}_{(\cal T)}} {e^\nu \left( {1-\omega {{d\phi } \over {dt}}} 
\right)}\,dt \, , \EQN Landau;c\cr} $$
where the integral over $dR$ is performed at $z=f(R)$  and that  over $dt$ is 
performed along a given trajectory ($\cal T$),
the line element on the disk  \Ep{diskmetric}
 reads 
$$d\,\sigma ^2=-{\slashchar d }\tau _*^2+d\tilde R^2+R_c^2d\phi ^2 . \EQN linedisk$$
For circular flows
($ {\slashchar d }\tau_*= {e^\nu [ {1-\omega {{d\phi } / {dt}}} ]}\,dt$),   
$V$ may be re-written   as
$$ V = R_c \,{ d  \phi \over {\slashchar d } \tau_*}. \EQN Vphi$$ 
 Equation \Ep{ Vphi} is  inverted to yield the angular velocity of the 
flow as measured at infinity:
$$\Omega \equiv {{d\phi } \over {dt}}={{V e^\nu /R_c} \over {1+\omega V e^\nu 
/R_c}} .\EQN OMEGA$$
Similarly,  the covariant specific angular momentum $h$ of a given 
particle reads 
 in terms of these variables
$$h\equiv {{p_\phi } \over \mu }=\gamma_{t\,\phi }u^t+\gamma_{\phi \,\phi 
}u^\phi =
{1 \over {\sqrt {1-V ^2}}}\left( {R_c V +\omega \,e^{\nu }} \right)\, , \EQN H$$
where $\mu$ is the rest mass of the particle.
The covariant specific energy $\epsilon$ of that particle reads 
$$\epsilon \equiv -{{p_t } \over \mu }=-\gamma_{t\,\phi }u^\phi-\gamma_{t \,t }u^t 
=
{e^{\nu }  \over {\sqrt {1-V ^2}}} \, .\EQN epsilon$$
which gives the central redshift, 
$1+z_c= \exp{(-\nu) }$.
The velocity of zero angular momentum observers (so called 
ZAMOs) follows from \Eq{H}:
$$ V_z = -{\omega e^{\nu} \over R_c}\, . \EQN ZAMO$$
With these observers, the cuts $z=f(R)$ may be  extended   inside ergoregions 
where the dragging of inertial frames induces apparent superluminous motions  as measured 
by locally static observers. 
The circumferential radius  $R_z$  as measured by ZAMOs is  Lorentz contracted  with 
respect to $R_c$, becoming 
$$R_z\equiv \sqrt{g_{\phi \phi }}= R_c \left[1-V_z^2 \right]^{1/2} , \EQN RZAMO$$
while the velocity flow measured by ZAMOs is
$$ V_{/z}= { V-V_z \over 1- V V_z } \, . \EQN VZAMO$$

The total angular momentum of the disk $H$ follows from the 
asymptotic behaviour of the vacuum field $\omega$ at large radii, 
$ \omega \rightarrow  -2 H/r  $.
Define the binding energy of the disk $\Delta E$ as the difference between
 the baryonic rest mass 
$M_0$ and the total mass-energy of the disk $M$ as measured from infinity  given 
by
$ \nu \rightarrow -M/r$:
$$\Delta E = M_0 - M. \EQN binding$$
$M$ is also given by the Tolman formula, but corresponds, by construction, to  the 
mass of 
the line source for the vacuum field.
The baryonic rest mass can be expressed as 
$$ M_0 \equiv  \int \Sigma \, \sqrt{-g} \, U^\mu d S_\mu   = 
  \int {\left[  1 -\omega \Omega\right]^{-1} \over \sqrt{1-V^2} }  
\,\Sigma  \,2 \pi R_c d{\tilde R}\, ,  \EQN baryonic $$
where  the baryonic energy density $\Sigma$ can be related to the energy
density $\varepsilon$ through some yet unspecified equation of state or the 
knowledge of a distribution function.
\subsection{ Constraints on the internal solutions }

The choice of the profile $z =f(R)$ 
is open, provided that it leads to meaningful physical quantities.
Indeed it must lead to solutions which satisfy
$\varepsilon \geq 0,\, p_{ {}_R}\geq 0, \, p_\phi \geq 0, \, {\rm and} \, V \leq1$
while the dominant energy condition implies also that 
$\varepsilon \geq p_\phi  \,{\rm and } \quad 
\varepsilon \geq p_{ {}_R}  $. 
These translate into the following constraints on $f$ 
$$\EQNdoublealign{  
&
 \varepsilon \geq 0 \, \, & \Rightarrow \, \, 
\sqrt {{\cal N}^2-{\cal O}^2}+({\cal N}-2{\cal Z})\geq 0\, ;  \EQN  constraint;a
\cr & 
0\leq V\leq 1 \, \, &\Rightarrow 
\, \, {\cal N}\geq {\cal O} \, \, {\rm and } 
\, \, 
   {\cal O}\geq 0 \, ;  \EQN  constraint;b
\cr 
&
0 \leq p_\phi \leq \varepsilon \, \, & \Rightarrow \, \,
4{\cal Z}\,{\cal N}\geq ({\cal O}^2+ 4{\cal Z}^2) \, \, {\rm and } \, \,
{\cal N}\geq 2{\cal Z} \, ;
\EQN  constraint;c \cr 
& 
0
\leq p_{ {}_R} \leq \varepsilon \, \, &\Rightarrow \, \,  
f'\leq0 \, \, {\rm and } 
\, \,
-2 f'/R \leq \sqrt {{\cal N}^2-{\cal O}^2}+({\cal N}-2{\cal Z})\, ;  \EQN  constraint;d
\cr
} 
$$
given 
$${\cal N}\equiv {{f'} \over R}+2{{\partial \nu } \over {\partial
N}}\, ,\quad 
  {\cal O}\equiv{{e^{2\nu }} \over R} 
{{\partial \omega } \over {\partial N}}\,,\quad {\rm and } \quad  {\cal Z}\equiv\left( {{{\partial \zeta }
 \over {\partial N}}-{{f''} \over {1+f'^2}}} \right) \, . \EQN NOZ $$

The existence of a solution satisfying this set of constraints 
can be demonstrated as follows:
in the limit of zero pressure and counter-rotation ({\it i.e.} $ {\cal O }\rightarrow 0,
{\cal N }\rightarrow 2 \partial \nu / \partial z, \, {\rm and} \quad {\cal Z }\rightarrow
 \partial \zeta / \partial z$), any cut $z= {\rm Const }\equiv b \gg 
m$ satisfies
\Eqs{ constraint}. By continuity,  there exist solutions with proper 
rotation
and partial pressure support.
In practise, all solutions given in sections~6 and~7 satisfy the
constraints
\Ep{ constraint}.
Note that in the limit of zero radial pressure, \Eq{ constraint;b} implies  $
{\partial } / {\partial z} \left[ \omega/R + \exp 2\nu \right]\leq0$.
\subsection{ Ansatz  for the profile }
In the following discussion, the section $z=f(R)$ is  chosen so that the corresponding radial 
pressure gradient balances 
a given fraction  of the gravitational  field which would have  occurred had 
there been no radial pressure within the disk. This choice `bootstraps' 
calculations for all relevant physical quantities in terms of a single 
degree of freedom ({\it i.e.} this fraction $\eta$), rather than a complete 
functional. 
This gives
$$  { \partial p_{ {}_R} \over \partial {\tilde R}}
=
-  {\eta \over (1-V^2)} { \partial \nu \over \partial {\tilde R} } \left[ \varepsilon + p_{ 
{}_R} + V^2( p_\phi-p_{ {}_R})
\right].  \EQN Ansatz $$
On the r.h.s. of \Eq{Ansatz},
 $p_{ {}_R}$ is  put to zero  and  $V,p_\phi,\varepsilon$ are re-expressed in 
terms of $\zeta, \nu,\omega$ via \Eqs{FinalV} with $f'= f''=0$ and $z=\rm Const$.
On the l.h.s. of \Eq{Ansatz},  $p_{ {}_R}$  is chosen according to 
\Eq{FinalV;d}.
In practise, this Ansatz for $f$ is a convenient way to investigate a
parameter space which is likely to be stable with respect to ring
formation as discussed in the next subsection.  In principle, a cut,
$f$, could be chosen so as to provide a closed bounded symmetrical
surface containing all fictitious sources. Indeed, by symmetry no flux
then crosses the $z=0$ plane beyond the cut, and the energy
distribution is therefore bound to the edge of that surface: this
corresponds to a finite pressure supported disk.  It turns out that in
general this choice is not compatible with all the constraints
enumerated in the previous section. More specifically, the positivity
of $p_\phi$ fails for all finite disk models constructed.  The Ansatz
described above gives for instance an upper bound on the height of
that surface when assuming that all the support is provided by radial
pressure.  This height is in turn not compatible with positive
azimuthal pressure at all radii.

\subsection{ Stability }

To what extent are the equilibria studied in the previous sections likely 
to be stable under the action of disturbances?
The basic instabilities  can be categorised as follows:

dynamical instabilities, which are intrinsic to the dynamical parameters of the disks,
 grow on an orbital time scale, and typically have drastic 
effects on the structure of the system. 

secular instabilities, which arise owing to dissipative
mechanisms such as viscosity or gravitational radiation, grow on a
time scale which depends on the strength of the dissipative mechanism involved,  
and slowly drive 
 the system along a sequence of dynamical equilibria.

Amongst dynamical instabilities, kinematical instabilities correspond
to the instability of circular orbits to small perturbations, and
collective instabilities arises from the formation of growing waves
triggered by the self-gravity of the disk.  Rings, for instance, will
be generated spontaneously in the disk if the local radial pressure is
insufficient to counteract the self-gravity of small density
enhancements.  Even dynamically stable non-axisymmetric modes may
drive the system away from its equilibrium by radiation of
gravitational waves which will slowly remove angular momentum from the
disk.  For gaseous disks, viscosity and photon pressure will affect
the equilibria.  Radiative emission may disrupt or broaden the disk if
the radiation pressure exceeds the Eddington limit.  The energy loss
by viscosity may induce a radial flow in the disk.  However, the disks
discussed in this paper have anisotropic pressures inappropriate for
gaseous disks and the accurate description of the latter two processes
requires some prescription for the dissipative processes in the gas.
The scope of this section is therefore restricted to a simple analysis
of the dynamical instabilities.

Turning briefly to the corresponding Newtonian  problem,
 Toomre (1963)\cite{Toomre} gave the local criterion for radial collective instability of stellar disks,
$$  \sigma_{{}_R} \geq { 3.36  G \Sigma_0  \over \kappa}, \EQN Toomre$$ 
where $\Sigma_0$ is the local surface density, $\sigma^2_R$ the radial velocity 
dispersion,
and $\kappa$ the epicyclic frequency of the  unperturbed stars. 
This criterion is derived from the first critical  growing mode of  the
 dispersion relation for  radial waves. 
The corresponding critical wavelength is  Jeans length $\sim \ 2\sigma_R^2 /G 
\Sigma_0  $.

For the relativistic disks described in this paper, spacetime is
 locally flat, which suggests a direct translation of \Eq{ Toomre}
 term by term.  The constraints that stability against ring formation
 places on these models will then be addressed at least qualitatively
 via the Newtonian approach.  The proper relativistic analysis is left
 to further investigation.  The relativistic surface density
 generalising $\Sigma_0$ is taken to be the co-moving energy density
 $\varepsilon$ given by \Eq{ FinalV;b}.  The radial velocity
 dispersion $\sigma_{{}_R}^2$ is approximated by $p_{
 {}_R}/\varepsilon$.  The epicyclic frequency is calculated in the
 appendix.  Putting \Eqs{ kappa} and \Ep{LNZ} into \Eq{ Toomre} give
 another constraint on $f$ for local radial stability of these disks.
 Note that the kinematical stability of circular orbits follows from
 \Eq{ radialEQ } by requiring $\kappa^2$ to be positive.

\section{Stellar dynamical equilibria for rotating super-massive disks}

The method described in section 2 will generally induce rotating disks with
anisotropic pressures ($p_\phi \neq p_R$). These objects may 
therefore be  described in the context of stellar dynamics.  It should
be then  checked that there exist a stellar equilibria compatible with a given
vacuum field and a given cut $z=f(R)$. 
\subsection{Distribution functions for  rotating super-massive disks}
Here a general procedure to derive all distribution functions
corresponding to a specified stress energy tensor is presented for
disks with non-zero mean rotation (a more direct derivation for
counter-rotating disk is given in Appendix~B). The detailed
description of the dynamics of the disk follows.  
\nl In Vierbein components, the stress energy tensor reads
$$T^{(\alpha \beta )}=\int {\int {{{f_{\star}(\epsilon,h)\,P^{(\alpha
 )\,}P^{(\beta )}\,} \over {P^{(t)}}}}}\,dP^{(R)}dP^{(\phi )} \, ,
 \EQN tv$$ where $f_{\star}(\epsilon,h)$ is the distribution of stars at
 position $R,\phi$ with momentum $ P^{(R)},P^{(\phi )}$. For a
 stationary disk, it is a function of the invariant of the motion
 $\epsilon,h$.  Now for the line element \Ep{diskmetric}:
$$ \eqalign{ d\sigma^2 = &
  - {e^{2\,\nu }}\,{{{\left(\it dt -\omega \, d\phi\right)}}^2}  
 +{e^{ 2\,\zeta-2\,\nu   }}\,(1+{f'}^2)\, {{{\it
  dR}}^2}  + {{}{R^2}{{e^{-2\,\nu }}}\, d \phi^2},  \cr}
\EQN WeylPmetric
  $$ the Vierbein momenta read
$$\EQNalign{
P^{(\phi)} &=  e^{\nu} \,R^{-1} (h- \omega \, \epsilon) \, ,\EQN Pv;a \cr
P^{(t)} &= e^{-\nu}\, \epsilon \, , \EQN Pv;b \cr
P^{(R)} &= \left[ \epsilon^2 e^{-2 \nu} - e^{2\nu} \, R^{-2}
 \, (h- \omega \, \epsilon)^2 -1 \right]^{1/2} \, . \EQN Pv;c \cr} $$
Calling $\chi = h/\epsilon \,\, {\rm and } \,\, \vartheta = 1/\epsilon^2$, \Eqs{Pv} becomes
$$\EQNalign{
P^{(\phi)} &=  e^{\nu} \,R^{-1}\, \vartheta^{-1/2} \,  ( \chi- \omega ) \, ,\EQN Pveta;a \cr
P^{(t)} &= e^{-\nu}\, \vartheta^{-1/2} \, , \EQN Pveta;b \cr
P^{(R)} &= \vartheta^{-1/2} \, \left[  e^{-2 \nu} - e^{2\nu} \, R^{-2}
 \, (\chi- \omega )^2 -\vartheta \right]^{1/2} \, . \EQN  Pveta;c \cr} $$
In terms of these new variables, the integral element $d P^{(\phi)} d P^{(R)}$ becomes
$$ d P^{(\phi)} d P^{(R)} = \left|{ \partial P^{(\phi)} \,\partial P^{(R)}
 \over \partial  \chi \,\, \partial\vartheta } \right| d \chi \,d \vartheta
=  { e^{-\nu} R^{-1}  \, \vartheta^{-2} \, d \chi \,d \vartheta \over
 2 \sqrt{ e^{-2 \nu} - e^{2\nu} \, R^{-2}
 \, (\chi- \omega )^2 -\vartheta } }\, . \EQN dJacob$$
Given \Eq{dJacob}  and \Eq{Pv;b}, \Eq{tv} may be rewritten as 
$$T^{(\alpha \beta )}=\int \int P^{(\alpha )}P^{(\beta )} 
 \, { f_{\star}(\epsilon,h)\, R^{-1}  \, \vartheta^{-3/2} \, d \chi \,d \vartheta \over
 2 \sqrt{ e^{-2 \nu} - e^{2\nu} \, R^{-2}
 \, (\chi- \omega )^2 -\vartheta } } \, . \EQN Tfnew$$
In particular,
$$
\EQNalign{
R \,T^{( t t )}\, e^{2 \nu} &=\int \int 
 \, { f_{\star}(\epsilon,h) \, \vartheta^{-5/2} \, d \chi \,d \vartheta \over
 2 \sqrt{ e^{-2 \nu} - e^{2\nu} \, R^{-2}
 \, (\chi- \omega )^2 -\vartheta } } \, , \EQN T11;a \cr
R^2 \, T^{( t \phi )}&=\int \int  (\chi -\omega)
 \, { f_{\star}(\epsilon,h)  \, \vartheta^{-5/2} \, d \chi \,d \vartheta \over
 2 \sqrt{ e^{-2 \nu} - e^{2\nu} \, R^{-2}
 \, (\chi- \omega )^2 -\vartheta } } \, , \EQN T11;b \cr
R^3 \, T^{(\phi \phi )} \, e^{-2\nu} &=\int \int (\chi -\omega)^2
 \, { f_{\star}(\epsilon,h)  \, \vartheta^{-5/2} \, d \chi \,d \vartheta \over
 2 \sqrt{ e^{-2 \nu} - e^{2\nu} \, R^{-2}
 \, (\chi- \omega )^2 -\vartheta } } \, . \EQN T11;c \cr
}
$$
 Note that given \Eqs{T11}, $T^{( R R )}$ follows from the equation of 
radial support. Note also that
$$
R_c(T^{( t t )}- T^{( \phi \phi )}-T^{( R R )}) =\int \int 
 \, { f_{\star}(\epsilon,h)\, e^{-\nu} \, \vartheta^{-3/2} \, d \chi \,d \vartheta \over
 2 \sqrt{ e^{-2 \nu} - e^{2\nu} \, R^{-2}
 \, (\chi- \omega )^2 -\vartheta }}\, , \EQN $$ 
or equivalently,
$$T^{( t t )}- T^{( \phi \phi )}-T^{( R R )}
 = \int \int{f_{\star}(\epsilon,h)\over P^{(t)} } 
\, d P^{(\phi)}\, d P^{(R)}  \, . \EQN tolman
$$
Equation \Ep{tolman} yields the total gravitating mass. It is known as
Tolman's formulae and can be derived directly from  the 
relation $P^\mu P_\mu =-1$ and \Eq{tv}.
 \nl
Let
$$ F(\chi,R)= {1\over 2} \int
 { f_{\star}(\epsilon,h) \vartheta^{-5/2} \,d \vartheta \over
  \sqrt{ e^{-2 \nu} - e^{2\nu} \, R^{-2}
 \, (\chi- \omega )^2 -\vartheta } } = \int\limits_1^{\cal Y}
  { f_{*}(\chi,\vartheta) \,d \vartheta \over
 2 \sqrt{ {\cal Y} -\vartheta } } \, , \EQN Fchi$$
 $${\rm where} \quad  f_{*}(\chi,\vartheta) = f_{\star}(\epsilon,h) \vartheta^{-5/2}\, ,\quad {\rm  and} \quad 
 {\cal Y}(R,\chi)=
 e^{-2 \nu} - e^{2\nu} \, R^{-2} \, (\chi- \omega )^2
\, .\EQN eqcalY $$
 The set of  \Eqs{T11} becomes
$$
\EQNdoublealign{
R \,T^{( t t )}\, e^{2 \nu} &=\int F(\chi,R)\, d \chi \,  &=R e^{-2 \nu} \int {\bar F}( v_\phi,R)\, d  v_\phi \, ,  \EQN T12;a
\cr R^2 \, T^{( t \phi )}&= \, \int F(\chi,R) \, (\chi -\omega)\, d \chi
\, &= R^2  e^{-4 \nu} \int v_\phi \, {\bar F}(v_\phi ,R)\, d v_\phi \, ,  \EQN T12;b \cr R^3 \, T^{(\phi \phi )} \, e^{-2\nu} &=\int
F(\chi,R)\, (\chi-\omega)^2\, d \chi \, &=R^3 e^{-6 \nu} \int v_\phi^2 \, {\bar F}( v_\phi ,R)\, d v_\phi \,  .\EQN T12;c \cr } $$ 

 The {\it r.h.s} of
\Eqs{T12} was re-expressed convinienlty in terms of the new variable
$ v_\phi \equiv e^{\nu} (\chi - \omega)/R_c $, and the new function $
{\bar F}(v_\phi,R) \equiv F(\chi = R_c v_\phi e^{-\nu} +
\omega  ,R) $ 
in order to stress the analogy with the classical identities.  Note
that in the limit of zero pressures, $v_\phi$ reduces to $V$, the
geodesic velocity of stars on circular orbits.  In \Eq{T12}, the
dragging of inertial frames requires to fix simultaneously three
moments of the velocity distribution to account for the given energy
density, pressures and rotation law. This third constraint
 does not hold  in the corresponding  Newtonian
problem, nor  for static relativistic disks as described in
appendix~B.
Any positive function $F(\chi,R)$ satisfying the moment constraints
\Eqs{T12} corresponds to a possible choice. For instance the parametrisation
is carried in  Appendix~C for $\bar F$ distributions corresponding to
powers of Lorentzians in azimuthal momenta. 
 Let ${\tilde
F}(\chi,{\cal Y}) = F(\chi,R)$,\,\, where ${\cal Y}(R,\chi)$ is given
by \Eq{eqcalY}.  Written in terms of $\tilde F$, the integral equation
\Eq{Fchi} is solved for $ f_{*}$ by an Abel transform
     $$ f_{*}(\chi,\vartheta)= {2\over \pi
 }\int\limits_{1}^{\vartheta}
\left({\partial {\tilde F}\over \partial {\cal Y} } \right)_{_{_\chi}}
 \, { \,d {\cal Y} \over \sqrt{ \vartheta -{\cal Y} } } \, . \EQN $$
  The latter integration may be carried in $R$ space and yields $$
  f_{\star}(\epsilon,h) = {2 \over \epsilon^4 \pi}  \int \limits_{{\cal R}_e}^{{\cal R}_p} 
\left({\partial { F}\over \partial {R} } \right)_{_{_\chi}}
 \, { \,d {R} \over
  \sqrt{-\left( {P^{(R)}}\right)^2 }}
-{2 \over \epsilon^4 \pi}  \int \limits_{{\cal R}_a}^{\infty} 
\left({\partial { F}\over \partial {R} } \right)_{_{_\chi}}
 \, { \,d {R} \over
  \sqrt{-\left( {P^{(R)}}\right)^2 }} \, , \EQN finaldf $$ where $P^{(R)}$ is given by
\Eq{Pv;c} as a function of $R,\epsilon$, and $h$. 
Here $F$ 
is chosen  to satisfy
\Eq{T12} which specifies the stress energy components $T^{(\alpha \beta )}$.
The integration limits correspond to two branches: the first branch
is to be carried between ${\cal R}_e(h)$, the inner radius of a star on
a ``parabolic''  orbit with momentum $h$, and ${\cal
R}_p(h,\epsilon)$, the perigee of a star with invariants $(h,\epsilon)$;
the second branch contributes negatively to \Eq{finaldf} 
and corresponds to radii larger than ${\cal R}_a(h,\epsilon)$,
the apogee  of a star with invariants $(h,\epsilon)$. Note 
that the derivative  in \Eq{finaldf} is taken at constant reduced
momentum $\chi = h/\epsilon$. This inversion procedure is 
illustrated on \Fig{kerrDF}.
 All properties of the flow 
follow in turn from $f_{\star}(\epsilon,h)$. For instance,
the rest mass surface density, $\Sigma$,
 may be evaluated from the  detailed ``microscopic''
behaviour of the stars, 
leading to  an estimate of the binding energy of the stellar cluster.

\figure{YR}
\Caption
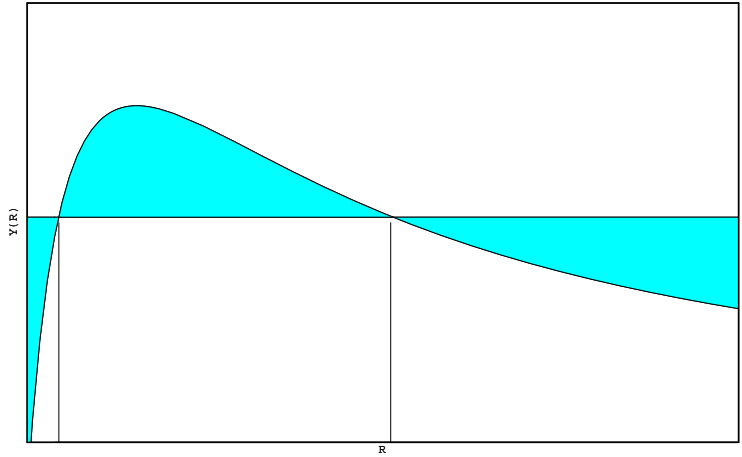
sketch for the effective potential $Y(R)$ as a function of radius.
The horizontal line corresponds to the energy level $1/\epsilon^2$,
the absisse line corresponds to the escape energy $1/\epsilon^2=1
$.  The ordinate axis corresponds to ${\cal R}_e(h,\epsilon)$, the
inner radius of a star on a ``parabolic'' orbit with momentum $h$, the
left vertical line corresponds to ${\cal R}_p(h,\epsilon)$, the
perigee of a star with invariants $(h,\epsilon)$; the right
vertical line corresponds to ${\cal R}_a(h,\epsilon)$, the apogee
of a star with invariants $(h,\epsilon)$.  The area between the
curve $Y=Y(R)$ and the lines $Y= 1/\epsilon^2$ and $R={\cal R}_e$
are shaded, defining three regions from left to right. The middle region
does not contribute to \Eq{finaldf}.  The equation $Y=
1/\epsilon^2$ has two roots corresponding to ${\cal R}_p(h,\epsilon)$  and  ${\cal R}_a(h,\epsilon)$, while $Y= 1$ has two roots corresponding to
infinity and ${\cal R}_e$. The sign of the slope of $Y(R)$ gives
the sign of the contribution for each branch in \Eq{finaldf}.
\endCaption
\endfigure 
\subsection{A simple equation of state}
An equation of state for these rotating disks with planar anisotropic
pressure tensors is alternatively found directly while assuming
(somewhat arbitrarily) that the fluid corresponds to the superposition
of two isotropic flows going in opposite directions.  In other words,
the anisotropy of the pressures measured in the frame co-moving with
the mean flow $V$ is itself induced by the counter rotation of two
isotropic streams.  (These counter rotating streams are described in
more details in the next section.)  In each stream, it is assumed that
the pressure is isotropic and that the energy is exchanged
adiabatically between each volume element. 
 The detailed derivation
of the corresponding relationship between the surface density and the
pressure of each stream is given in Appendix~A. 
The above set of assumption yields the following prescription
 for $\Sigma$ :
$$\Sigma = {{ \varepsilon - p_\phi -p_{_R} +
\sqrt{(\varepsilon - p_\phi -p_{_R})(\varepsilon - p_\phi +3 p_{_R})}  }
\over 2  \sqrt{1-V_0^2}},\EQN Sigma $$ where $V_0= 
\sqrt{(p_\phi - p_{_R})/(\varepsilon +  p_{_R})}$ is the 
counter rotating velocity measured in the  frame co-moving with $V$ which induces 
$p_\phi \neq p_{_R}$. In the classical limit,  $\Sigma \rightarrow 
(\varepsilon - p_\phi) /  \sqrt{1-V_0^2}$. The binding energy of these rotating disks is computed
from 
\Eq{Sigma } together with \Eqs{baryonic} and  \Ep{FinalV}.

\section{ Application: warm counter-rotating disks  }

For simplicity, warm counter-rotating solutions are presented first,
avoiding the non-linearities induced by the dragging of inertial
frames.
  These solutions generalise those of BLK, while
implementing partial pressure support within the disk.  Formally this
is achieved by putting $\omega(R,z)$ identically to zero in
\Eq{WeylPmetric}; the metric for the axisymmetric static vacuum
solutions given by Weyl is then recovered: $$ ds^2 = - {e^{2\,\nu
}}\,{{{\it dt}}^2} +{e^{ 2\,\zeta -2\,\nu }}\,( {{{\it dR}}^2} +{{{\it
dz}}^2}) + {{{}{R^2}\,{{e^{-2\,\nu }}} d \phi^2}},
\EQN Weylmetric
  $$ 
where the functions $\zeta(R,z)$ and $\nu(R,z)$ are generally of
the form (cf. Eq. (BLK 2.21)) $$\EQNalign{ & \nu=-\int {
W(b) \, db \over \sqrt{R^2+(|z|+b)^2 }}, \EQN nuzeta;a \cr & \zeta
=\int \int W(b_1) W(b_2) Z(b_1,b_2) \, d b_1\, d b_2,
\EQN nuzeta;b \cr }$$
with $Z$ given by  $$Z = - {1\over 2r_1r_2} \left\{1 - \left({r_2-r_1\over b_2-b_1}\right)^2 
\right\}  \geq 0  \, . \EQN  (BLK 2.33)$$
Here $b_1$ and $b_2$ are the distances of two points below the disk's center 
and $r_1$ and $r_2$ are the distances measured from these points to a point 
above the disk.  Below the disk, $Z$  is given by reflection with respect to the plane  
$z=0$.
When the line density of
fictitious sources is of the form $W(b) \propto b^{-m}$
(Kuzmin-Toomre), or $W(b) \propto { \delta^{(m)}\,(b) }$
(Kalnajs-Mestel), the corresponding functions $\zeta,\nu$ have
explicitly been given in B\`\i c\'ak, Lynden-Bell \& Pichon (1994)\cite{BLP}.
  For the metric of \Eq{
Weylmetric}, the Lanczos tensor \Ep{Lanczos} reads $$\left[
{\matrix{{\cal L }^{(t)(t)}\cr {\cal L }^{(R)(R)}\cr {\cal L
}^{(\phi)(\phi)}\cr }} \right]={{e^{\nu -\zeta }} \over {4\pi \sqrt
{1+f'^2}}}\left[ {\matrix{{{{f''} \over { {1+f'^2}}}+{{f'} \over
R}+{{\partial (2\nu -\zeta )} \over {\partial N}}}\cr {{{-f'} \over
R}}\cr {{{-f''} \over { {1+f'^2}}}+{{\partial \zeta } \over {\partial
N}}}\cr }} \right]. \EQN FinalC $$

Consider again counter-rotating disks made of two equal streams of
stars circulating in opposite directions around the disk center.  The
stress energy tensor is then the sum of the stress energy tensor of
each stream: $$ \eqalign{
\tau ^{(a)(b)} & =  \left[ {\matrix{{\varepsilon}&0&0\cr
0&{p_{ {}_R}}&0\cr
{0 }&0&{p_\phi}\cr
}} \right] 
=
{2 \over {1-V_0^2}}\left[ {\matrix{{\varepsilon_0 +p_0 \,V_0 ^2}&0&0\cr
0&{(1-V_0^2)\,p_0}&0\cr
{0 }&0&{p_0 +\varepsilon_0 V_0 ^2}\cr
}} \right] ,
 \cr } \EQN stressC $$
  the pressure $p_0$ in each stream being chosen to be isotropic in the plane.
Subscript $(\quad)_0$ represents quantities measured for one stream. Expressions for 
$\varepsilon$,$\,p_\phi$ and $p_{ {}_R}$ follow from identifying  \Eqs{stressC} 
 and \Ep{FinalC} according
to  \Eq{T2K}.
Solving for $\varepsilon_0, \,V_0, \,{\rm and} \, p_0 $ in \Eq{stressC}, given 
\Eq{FinalC} and \Ep{T2K}
 gives 
$$\EQNalign{&\varepsilon_0 ={{e^{\nu-\zeta }} \over {4\pi \sqrt {1+f'^2}}}
\left[ {{{f''} \over {1+f'^2}}+{{\partial (\nu -\zeta )} \over {\partial N}}} \right], 
\EQN counter;a
\cr
  &p_0={{e^{\nu-\zeta }} \over {8\pi \sqrt {1+f'^2}}}\left[ {{{-f'} \over R}} \right],  
\EQN counter;b\cr
  &V_0^2 =  \left(  
{\partial \zeta  \over \partial N} - {f'' \over 1+f'^2} +{f' \over R} 
\right)
\left( 
{ f'' \over 1+f'^2}+{\partial (2\nu -\zeta ) \over \partial N }
\right)^{-1}\, .  \EQN counter;c\cr}$$
The above solution provides the most general counter rotating disk model with 
pressure support.
Indeed, any physical static disk will be characterised entirely by its vacuum field 
$\nu$
and its radial pressure profile $p_{ {}_R}$, which in turn defines  $W(b)$ 
and $f(R)$ uniquely according
to \Eqs{counter;b} and (BLK 2.33). The other properties of the disk are then 
readily
derived.

The angular frequency and angular momentum of these disks follow from
\Eq{H},
\Ep{OMEGA } on putting $\omega$ to zero and $V$ to $V_0$.
The epicyclic frequency $\kappa$ given by \Eq{kappa}
may be recast as   
$$ \kappa^2 = {e^{\nu} \over R_c^3}  { d h^2 \over d {\tilde R} }, \EQN 
$$ 
which relates  closely to the classical expression 
$\kappa^2 = R^{-3}  { d h^2 / d { R}} $.

The Ansatz given by \Eq{Ansatz} for $z=f(R)$ becomes, after the
substitutions $\varepsilon
\rightarrow \varepsilon_0, \, V \rightarrow V_0, \,p_{\phi} \rightarrow 0$, 
$$\EQNalign{  (f'/R)' &=  -\eta { \partial \nu \over \partial R } \left[ { \partial \nu \over \partial z }
- {1 \over 2} { \partial \zeta \over \partial z } \right]\, ,  \EQN choiceFC;a  \cr  
 &= \eta { \partial \nu \over \partial R }{ \partial \nu \over \partial z }
\left[ R { \partial \nu \over \partial R } -1 \right] \, , \EQN choiceFC ;b \cr }$$
 where $z$ is to be evaluated at $b$.
 Equation \Ep{ choiceFC;b} follows from \Eq{ choiceFC;a} given the $({}^R_z)$ component 
of the Einstein equation outside the disk.

The equation of state for a relativistic isentropic 2D flow of
counter-rotating identical particles is derived in appendix~A while
relating the perfect fluid stress energy tensor of the flow to the
most probable distribution function which maximises the Boltzmann
entropy.  It reads
  $$\varepsilon_0  =  2 \, p_0 + {\Sigma \over 1+ p_0/\Sigma } \, .  \EQN$$
Solving for $\Sigma$ gives
$$ {\Sigma} = {1\over 2}\left[{\varepsilon_0  - 2\,p_0 + 
      {\sqrt{\varepsilon_0  - 2\,p_0}}\,{\sqrt{\varepsilon_0  + 2\,p_0}}}\right] \, ,  \EQN sigma $$
which in the classical limit gives 
$\Sigma \rightarrow \varepsilon_0  - p_0 $.
 The binding energy of these counter rotating disks is computed here from 
\Eq{sigma } together with \Eqs{baryonic} and \Ep{counter}.
Alternatively, it could be computed by constructing 
 distribution functions using the inversion described  in  Appendix~B.
\section{ Examples of warm counter-rotating disks}
\subsection{Example 1 : The Self-similar disk }
The symmetry of the self-similar disk allows one  to reduce the partial 
differential equations 
corresponding to Einstein's field equations  to an 
ordinary differential equation with respect to the only free parameter $\theta 
=\arctan(R/z)$ (Lemos, 
89 \cite{Lemos}).
Lemos' solution may be recovered and corrected by the method presented
in sections~2 and~4.
  Weyl's metric in spherical polar co-ordinates
$(\rho,\chi,\phi)$ defined in terms of $(R,z,\phi)$ by $\rho = \sqrt(R^2+z^2)$ and 
$\cot(\chi) = z/R$ is
$$ d\sigma^2=-e^{2\nu }d\tau^2+\rho ^2\left[ 
e^{-2\nu }\,\sin^2(\chi )d\phi^2
 +e^{2\zeta -2\nu }\left( {{{d\rho } \over {\rho ^2}}^2+d\chi ^2} \right)
\right] .
 \EQN WeylSpher$$
A self-similarity argument led Lemos to the metric 
$$ds^2=-r^{2n}e^N dt^2+r^{2k}\left[ 
e^{2P-N}\,d\phi^2
+e^{Z-N}\left( {{{dr} \over {r^2}}^2+d\theta ^2} \right) 
\right],\EQN$$
where $P$, $N$, and $Z$ are given by
$$\EQNalign{P(\theta )&=\log \left[ {\sin (k+n)\,\theta } \right] 
,\EQN PLemos;a \cr
  N(\theta )&={{4n} \over {k+n}}\log \left[ {\cos (k+n)\,\theta /2} \right]
 ,\EQN PLemos;b \cr
  Z(\theta ) &=
 {{8n^2} \over {\left( {k+n} \right)^2}}  \log \left[ 
{\cos (k+n)\,\theta /2} \right] 
   + 2 \log(k+n).  
\EQN PLemos;c \cr} $$
Here the $\phi$'s  in both metrics have already been identified  since they
should  in both case vary uniformly in the range $[ 0, 2 \pi[$.
The rest of the identification between the two metrics follows from the 
transformation 
$$\EQNalign{&\chi =(n+k) \,\theta ,\EQN newvar;a \cr
  &\rho =r^{n+k}, \EQN newvar;b\cr
  &\tau =t, \EQN newvar;c \cr}  $$
with the result that $\nu$ and $\zeta$ take the form: 
$$\EQNalign{
  &2\nu =N+2n\log (r), \EQN 111;a   \cr
  &2\zeta =Z-2\log (n+k).\EQN 111;b \cr}  $$
This can be written in terms of the new variables  given in \Eqs{ newvar} as 
$$\EQNalign{ &\zeta ={{4n^2} \over {\left( {k+n} \right)^2}}\log ({\chi
  \over 2}),  \EQN NewPotential;a \cr &\nu ={{2n} \over {\left( {k+n} \right)^2}} 
\log ({\chi
  \over 2})+{n \over {k+n}} \log(\rho ). \EQN NewPotential;b \cr } $$
 The procedure described earlier may now be applied.
In order to preserve the self-similarity of the solution and match Lemos' solution, 
$f(R)$ is  chosen   so that  
 $$f'(R) = {\rm const.}=  \cot(\bar\eta) \quad {\rm say}. \EQN fprime$$
Then 
$${\partial  \over {\partial N}}\equiv \left( {{\partial  \over {\partial z}}} -f'
 {{\partial  \over {\partial R}}} \right)=
-{1 \over \rho \,\sin (\bar \eta )}{\partial  \over {\partial \bar \eta }} . \EQN LemosdN$$
Putting \Eqs{ NewPotential}  and \Eq{LemosdN} into \Eq{ FinalV} gives
$$\left[ {\matrix{{\varepsilon }\cr
{p_{ {}_R}}\cr
{p_{\phi }}\cr
}} \right]={r^{-k} \over {4\pi }}\left[ {\cos ({\bar \eta  \over 2})} 
\right]^{{\textstyle{{2n\left( {k-n} \right)} \over {(k+n)^2}}}}\left[
 {\matrix{{\cot (\bar \eta )+{{2n\,k} \over {\left( {k+n} \right)^2}} \tan(\eta /2)}\cr
{-\cot (\bar \eta )}\cr
{{{2n^2} \over {\left( {k+n} \right)^2}}\tan (\bar \eta /2)}\cr
}} \right],
  \EQN LemosLanczos $$
with $ \bar\eta $ to be evaluated at $(n+k) \, \pi /2 $.
The above equations  are in accordance with the solution found by Lemos
 by directly integrating the Einstein field equations.

\subsection{ Example 2: The Curzon disk }

\midfigure{Curzon}
\Caption
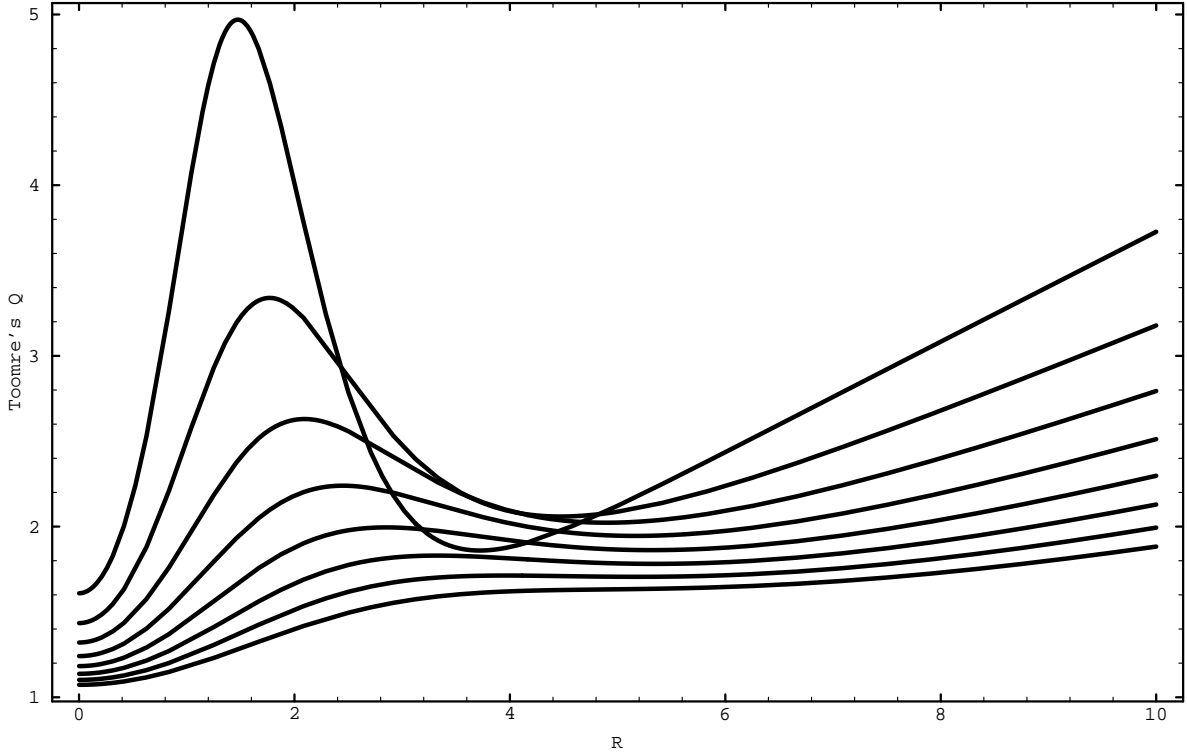
  the "Toomre Q number" $Q=  \sigma_{_R}\kappa / (3.36 \Sigma_0 )  $ 
is plotted here as a function of circumferential radius $R_c$ for different  compactness
parameter $\alpha =b/M$ when $\eta = 1/25$.
\endCaption
\endfigure 

The Curzon disk is characterised by the following pair of Weyl functions: 
$$\nu =-\alpha /r,\quad\quad\zeta =-\alpha ^2R^2/(2\,r^4), 
\EQN $$ 
with $ r=\sqrt{R^2+f^2} $,
given the dimensionless parameter 
$ \alpha= M/b  $ (recall that $G \equiv c \equiv 1$), all lengths being expressed in 
units of $b$. 
This disk is the  simplest example of the Kuzmin-Toomre family,
 where $ W(b) \propto \delta(b) $. It also corresponds to the building block of 
the expansion given in \Eqs{nuzeta}.
Equations \Ep{ FinalC} and \Ep{stressC} then imply  
\midfigure{Curzon2}
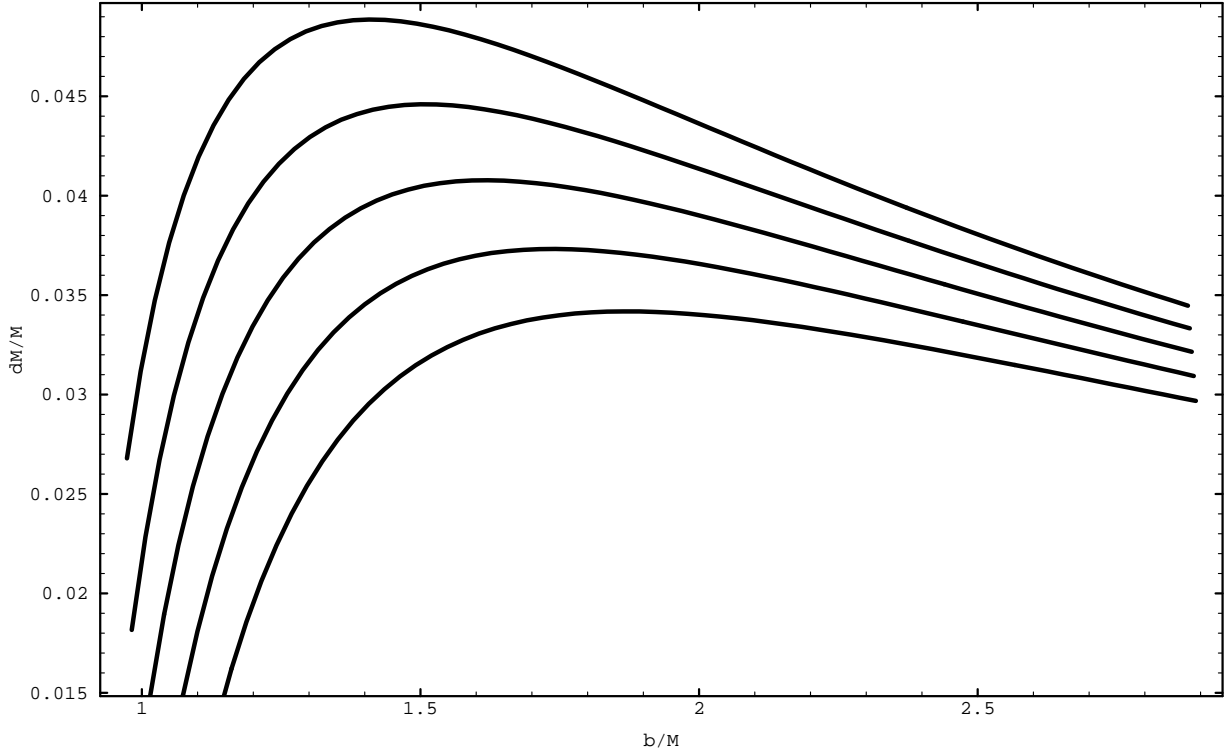
\Caption   the binding energy  over the  rest 
mass $\Delta E/M_0$ for Curzon disks, plotted against the compactness
parameter $\alpha= b/M$ for different  ratios $p_0/\varepsilon_0$ 
fixed by $\eta= 0.01,0.16, \cdots, 0.7$.
\endCaption
\endfigure
$$\left[ {\matrix{\varepsilon \cr
{p_{ {}_R} }\cr
{p_\phi}\cr
}} \right]={{e^{-{\alpha  \over r}\left( {1-{{\alpha \,R^2} \over {2r^3}}}
 \right)}} \over {4\pi b\sqrt {1+f'^2}}}\left[ {\matrix{{{{f''} \over {1+f'^2}}+{{f'} 
\over R}+
{\alpha  \over {r^6}}\left\{ {2f(r^3-\alpha \,R^2)+\left[ {\alpha \left( {R^2-f^2} 
\right)-2r^3} \right]Rf'} \right\}}\cr
{{{-f'} \over R}}\cr
{-{{f''} \over {1+f'^2}}+{{\alpha ^2} \over {r^6}}\left[ {2R^2f-\left( {R^2-f^2} 
\right)Rf'} \right]}\cr
}} \right].  \EQN curzonAll$$
 
The weak energy conditions $ \varepsilon \geq 0$, $\varepsilon +p_{
{}_R} \geq 0$, and $
  \varepsilon +p_\phi \geq 0$ which follow are in agreement with those found 
by Chamorow {\sl et al.}(1987)\cite{Chamorow}. 
Their solution was derived by direct integration of Einstein's 
static equations
using the harmonic properties of the supplementary unknown needed to avoid 
the patch
of Weyl metric above and below the disk. The cut $z=f(R)$ corresponds to the 
imaginary
part of the complex analytic function, the existence of which follows from the 
harmonicity of that new function.

The Ansatz \Ep{Ansatz} leads here to the cut
$$ f(R)= b\, + {\eta \over 2} \left[{ b \, m^2  \over 4 (b^2 + R^2) }-
{2 b  \, m^3 (4 b^2 + 7 R^2) \over 105 (b^2 + R^2)^{5/2} } \right]\, . 
\EQN CurzonCut$$
In \Fig{Curzon}, the ratio of the binding energy of these Curzon 
disks (which is derived from \Eqs{baryonic},
\Ep{sigma},\Ep{curzonAll} and \Ep{CurzonCut}) over the corresponding rest 
mass $M_0$ is plotted with respect to the compactness
parameter $b/M_0$ for different  ratios $p_0/\varepsilon_0$ measured at the origin.
The relative binding energy decreases in the most compact configurations because 
these contain too many unstable orbits ($\kappa^2 < 0$). A maximum ratio of 
about 5 $\%$ is reached.

\section{Examples of rotating disks and  an internal solution for the
Kerr metric  }
 The problem of constructing disks with  proper rotation and partial pressure support 
is more complicated but physically more 
appealing than that of counter rotating solutions.
 It is now illustrated on an internal solution for the Kerr
metric, and a generalisation of the solutions constructed by BLK and BLP
 to {\sl  warm rotating } solutions is sketched. 
\subsection{ A Kerr internal solution}
The  functions $(\nu,\zeta,\omega)$ for
 the Kerr metric in Weyl-Papapetrou form expressed in terms of spheroidal 
 co-ordinates ($R= s \sqrt{(x^2 \mp 1)(1-y^2)}, \, \,  z= s x y $)  are given by
$$e^{2 \nu}= {{-{m^2} + {s^2}\,{x^2} + {a^2}\,{y^2}}\over {{{\left( m + s\,x 
\right) }^2} +\
  {a^2}\,{y^2}}}, \EQN e2u $$
$$ e^{2 \zeta}={{-{m^2} + {s^2}\,{x^2} + {a^2}\,{y^2}}\over {{s^2}\,\left( {x^2} 
\mp {y^2}\
  \right) }}, \EQN e2k $$ and 
$$\omega = -{{2\,a\,m\,\left( m + s\,x \right) \,\left( 1 - {y^2} \right) }\over {-{m^2} 
+\
  {s^2}\,{x^2} + {a^2}\,{y^2}}}, \EQN omegaKerr $$
where $s = (\pm ( m^2-a^2))^{1/2}$.
Here $\pm$ corresponds to the choice of prolate $(+)$, or oblate $(-)$ spheroidal 
co-ordinates corresponding
to the cases $a<m$ and $a>m$ respectively.  In this set of co-ordinates, the prescription given in section~2 leads to completely algebraic solutions.
The normal derivative to the surface $z=f(R)$  reads (when $a>m$)
$$\left( {{\partial  / {\partial N}}} \right)  =
N(x,y)\left( {{\partial  / {\partial x}}} \right)+
N(y,x)\left( {{\partial  / {\partial y}}} \right), \EQN dNxy $$
where $$ N(x,y) = {{(x^2-1)\left[ {y/s+x\,\,(y^2-1)\,f'/R} \right]} 
/ {\left( {x^2-y^2} \right)}}. $$
Differentiating \Eqs{ e2u } through \Ep{ omegaKerr } with respect to $x$ and  $y$ together 
with
\Eq{ dNxy} leads via \Eqs{FinalV} to  all physical characteristics   of 
the Kerr disk in terms of the Kerr metric parameters $m$ and $a$, and
 the function $z=f(R)$ which must be chosen so as to provide relevant
 pressures and energy distributions according to \Eqs{ constraint}.
(B\`\i c\'ak \& Ledvinka (1993)\cite{Bicak} have constructed
independently a cold  Kerr solution when $a<m$.)
On the axis $R=0$, $f\equiv b, \,\, {\rm and } \, \,  f'' \equiv -c$, while \Eqs{ constraint}
imply
$ f'=0 \, {\rm at} \,  \, (x=b/s, \,\, y=1) $. 
This in turn implies  
$$\EQNalign{&\left(p_\phi\right)_0 =\left(p_{ {}_R}\right)_0={c \over {4\pi }}\sqrt 
{{{a^2+b^2-m^2} \over {a^2+(b+m)^2}}}, \EQN cond;a\cr
  & \eqalign{ \left(\varepsilon\right)_0 & = 
{1 \over {4\pi }}\sqrt {{{a^2+b^2-m^2} \over {a^2+(b+m)^2}}}  
\left[ {{ 2m \left[{\,(b+m)^2-a^2}\right] \over {\left( {a^2+(b+m)^2} \right)
\left( {a^2+b^2-m^2} \right)}}-c} \right]  \cr},
\EQN cond;b\cr}$$
where $(\quad)_0$ stands for (\quad) taken at $R=0$.
The constraint that all physical quantities should remain positive implies
$$\EQNalign{&b\geq \sqrt {m^2-a^2}, \EQN cond2;a \cr
  &c\leq 2m{{\,(b+m)^2-a^2} \over {\left( {a^2+(b+m)^2} \right)\left( 
{a^2+b^2-m^2} \right)}}. \EQN cond2;b\cr }$$
Equation \Ep{ cond2;a} is the obvious requirement that the surface of section should
not enter the horizon of the fictitious source. 
Similarly, ergoregions will arise when the cut $z=f(R)$ enter the torus $$({z\over m})^2
+({R\over m} -{a\over m}+ {1\over 2}{ m \over a})^2 = ({1\over 2} {m\over a})^2.  \EQN ergoregion $$
The analysis may be extended beyond the ergoregion via the ZAMO frame. The 
 characteristics of these frames are given by \Eq{ZAMO}, \Ep{RZAMO}  and 
\Ep{VZAMO}.
The central redshift $$1+z_c = e^{-\nu}={ { (m+b)^2 + a^2 }\over { b^2 + a^2 - m^2 
} }, \EQN redshift$$
can be made very large  for such models constructed when $b \rightarrow \sqrt{ m^2 
-a^2}$. 
\widetopfigure{KerrV}
\Caption
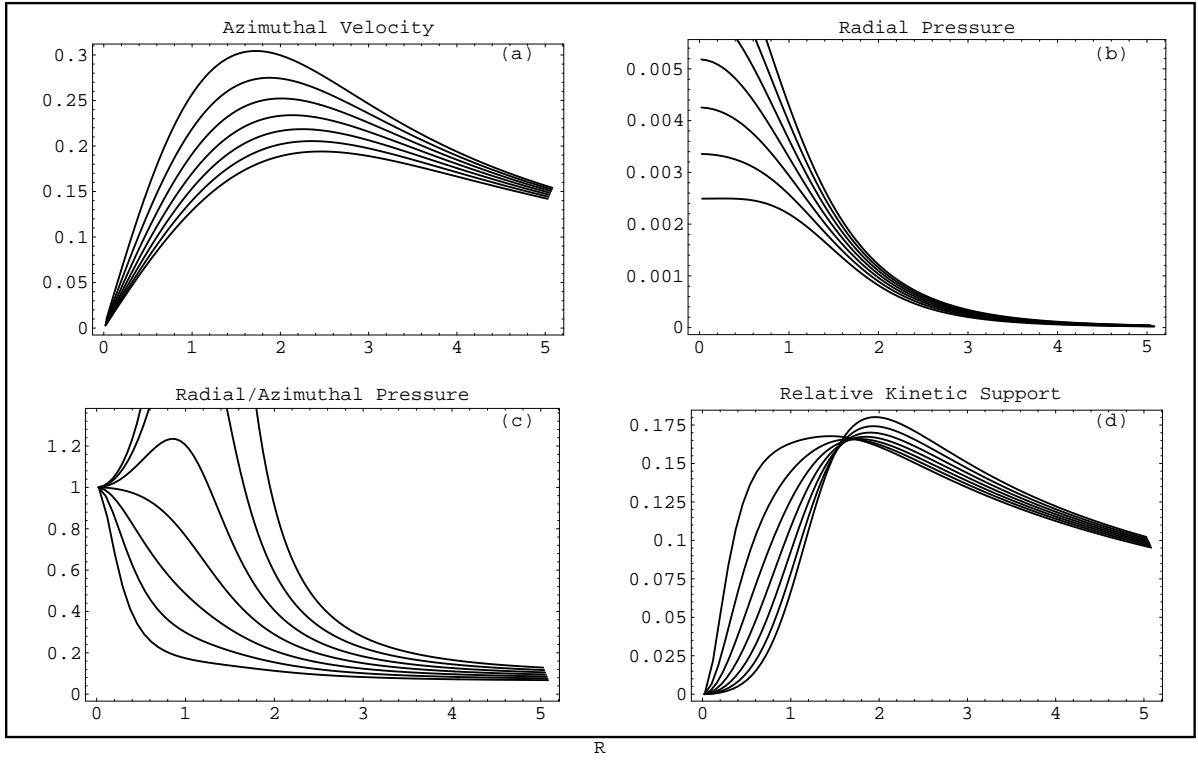
 the  azimuthal velocity (a), the radial pressure (b),
the ratio of radial over azimuthal pressure and 
the relative fraction of {Kinetic} Support (d) namely
$\varepsilon V^2 /( \varepsilon V^2+ p_\phi + p_{_R})  $
 for the $a/m = 0.5$  Kerr disk as a function of circumferential radius 
$R_c$,
for a class of solutions with decreasing  potential compactness,
$b/m = 1.1,1.2 \cdots 1.7$, and relative radial 
pressure support $ \eta/m =1/5 +2(b/m-1)/5 $ . 
\endCaption
\endfigure
\widetopfigure{KerrV2}
\Caption
 as in \Fig{KerrV} with $a/m=0.95$,  
$\eta/m= 1/15 + (b-1)/5$. Note the large relative kinetic support.
\endCaption
\endfigure
On \Fig{KerrV}, \, \Fig{KerrV2} and \Fig{KerrV3},   some characteristics of these  
Kerr disks with both $a > m$ and $a<m$ are illustrated.
 For simplicity, pressure  is implemented 
 via the cut \Ep{CurzonCut}.
 These  disks
  present anisotropy of the 
planar pressure tensor ($p_\phi \neq p_{ {}_R}$). 
The anisotropy follows from the properties of the $\omega$ vacuum field which gives rise to 
the circular velocity curve.  Indeed, in the outer part of 
the disk, the specific angular momentum $h$ tends asymptotically to a constant. 
The corresponding centrifugal force
is therefore insufficient to counter-balance the field generated by the $\nu$ function. 
As  this construction scheme generates such equilibria, the system compensates by
increasing its azimuthal pressure away from isotropy. 
When $a>m$, and $\eta =0$ the height of the critical
 cut corresponding to the last disk with positive pressures everywhere
scales like $a\, (a^2-m^2)^{1/2}$ in the range $2 \leq a/m  \leq 20$.
Note that this limit is  above that of the  highly relativistic motions 
which only occurs when $b$ tends asymptotically to the Kerr horizon $(a^2-m^2)^{1/2}$.
The binding energy which follows \Eq{Sigma} reaches values  as high as one tenth 
of the rest mass for the most compact configurations.
The inversion method described in section~4.1 was carried for the Kerr
disks and yields  distribution functions which characterise completely
the equilibria as illustrated on 
\Fig{kerrDF}. It was assumed that the velocity distribution of stars
at radius $R$ 
was a modified squared Lorentzian, as discussed 
in Appendix~C and shown on \Fig{Fphi}.
Note that disks with low rotation support present in their
outer parts two distinct maxima, one of which is counter rotating, in
agreement with the assumptions yielding the equation of state \Eq{Sigma}.
\figure{KerrV3}
\Caption
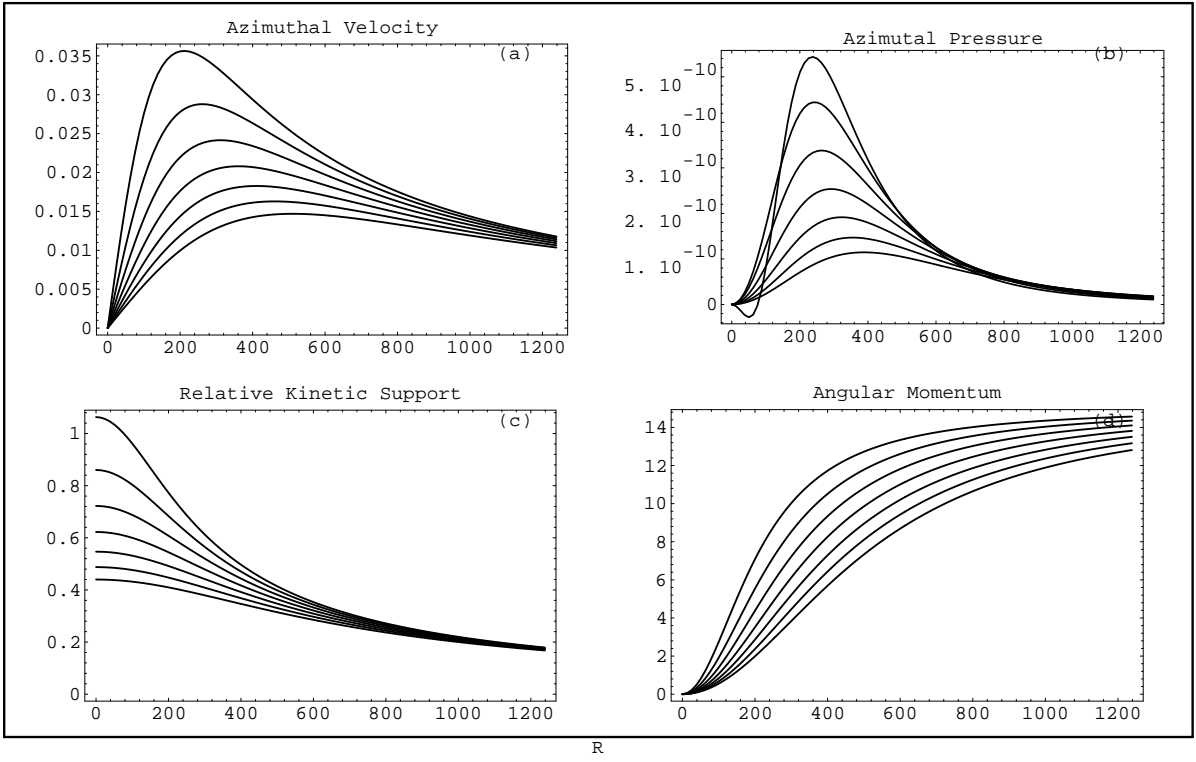
 the  azimuthal velocity (a), the azimuthal pressure (b),
the relative fraction of Kinetic Support (c), namely 
$\varepsilon V^2 /( \varepsilon V^2+ p_\phi )  $   and
 the angular momentum (d)
 for the $a/m=10$ zero radial pressure
 Kerr disk as a function of circumferential radius 
$R_c$,
for a class of solutions with decreasing  potential compactness
$b/m= 210,260 \cdots 510$. The first curve corresponds to a cut which induces
negative azimuthal pressure (tensions).  
\endCaption
\endfigure 
\figure{Fphi}
\Caption
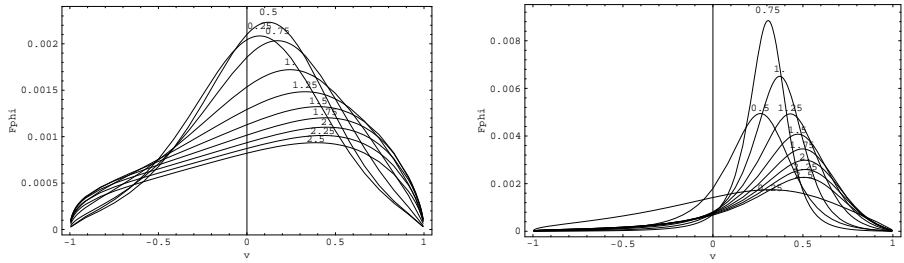
Parametrization of the number of star with 
velocity $v_\phi$ at radius $R$ 
as a squared Lorentzian (as discussed
in Appendix~C) for the Kerr disk. The corresponding radius is appended
on each curve.  The left panel correspond to $a=0.2$ while the right
panel corresponds to $a=0.8$.  The pressure law is given in the
caption of \Fig{kerrDF} where the corresponding distribution function
is illustrated. 
\endCaption
\endfigure 
\figure{kerrDF}
\Caption
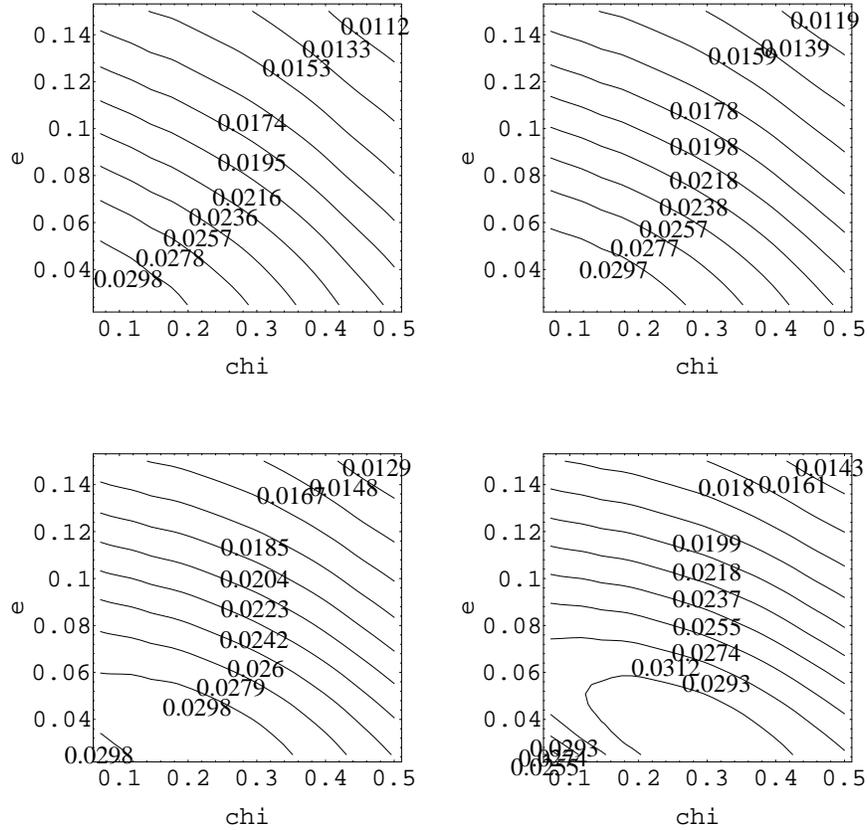
the isocontours of the distribution function for the Kerr disk as a
function of the reduced momentum $\chi = h/\epsilon$, and the relative eccentricity of
the orbit $\bar e_c = (\epsilon - \epsilon_\chi)/(1-
\epsilon_\chi)$ where $ \epsilon_\chi $ is the energy of
circular orbits with reduced momentum $\chi$.
The four panels correspond  to $a=0,0.1,0.2,0.3,0.4$ from left to right and
top to bottom,
while the pressure law was chosen to correspond to the cut 
$f(R)=-4(4 + 7R^2)/(525(1 + R^2)^{5/2}) + (16 + 15R^2)/(10(1 + R^2))
$ corresponding to a tepid disk.
The figures are centered on the inner parts of the disk to illustrate
the relative shift of  the maximum number of 
stars towards larger angular momentum for faster rotating disks.
The inversion assumed a squared Lorentzian for the distribution 
in angular velocity as discussed in  Appendix~C and illustrated on \Fig{Fphi}.
\endCaption
\endfigure 
\subsection{ Other rotating solutions}
 
The rotating disk models given in section~2 require prior knowledge of
the corresponding vacuum solutions. However, while studying the
symmetry group of the stationary axially symmetric Einstein Maxwell
equations, Hoenselaers, Kinnersley \& Xanthopoulos
(1979)\cite{HKX}\cite{HKX2} found a method of generating
systematically complete families of stationary Weyl-Papapetrou vacuum
metrics from known static Weyl solutions. Besides, the decomposition
of the Weyl potential $\nu_0$ into line densities provides a direct
and compact method of finding solutions to the static field.  The N+1
rank zero HKX transform is defined as follows: let $\nu_0$ and
$\zeta_0$ be the seed Weyl functions given by \Eqs{nuzeta}.  HKX
define the $N \times N$ matrix $\bf { \Gamma }$ parameterised by $ N $
twist parameters $\alpha_k$ and $N$ poles, $a_k$, $k=1, \cdots N$ as
$$\eqalign{&\left( {\Gamma^\pm } \right)_{k,k'}=
i\,\alpha_k\,{{e^{\beta_k}} \over {r_k}}\left[ {{{r_k-r_{k'}} \over
{a_k-a_{k'}}}\pm 1} \right]\cr }, \EQN gamma$$ with $r_k=r(a_k)=\sqrt
{R^2+\left( {z-a_k} \right)^2} \, $ and $ \beta_k=\beta (a_k) \, $,
$\beta$ being a function satisfying \Ep{beta;a} below.  Defining ${\bf
r} ={\rm Diag}(r_k)$, they introduce the auxiliary functions
$$\eqalign{&D^\pm =\left| {\bf{1}+ \bf{\Gamma^\pm} } \right|,
\,\,   {\rm and} \,\, L_\pm =
2D^\pm \,{\rm tr}\left[ {\left( {\bf{1}+{\bf \Gamma^\pm} } \right)^{-1}
{\bf\Gamma^\pm} \, {\bf r}}
 \right], \cr
  &\omega_\pm =D^+\pm e^{2\nu_0} D^- ,\,\,   {\rm and}\,\, { \cal M}_\pm =
\varpi\,\omega_\pm +2\left( {L_+\mp e^{2\nu _0} L_-} \right),
\cr}$$ 
which lead to the Weyl-Papapetrou potentials 
$$\EQNalign{e^{2\nu } &= e^{2\nu _0}\,\Re \left[ {{{D^-} \over {D^+}}} \right], 
\EQN WeylHKX;a\cr
  e^{2\zeta } &= k e^{2\zeta _0}\,\Re \left[ {{{D^-} {D^+}^*}} \right], \EQN 
WeylHKX;b\cr
  \omega = &2 \Im \left[ {{{{ \cal M}_+^*\left( {\omega _++\omega _-} \right)-{ 
\cal M}_-\left(
{\omega _++\omega _-} \right)^*} \over {\left| {\omega _+} \right|^2-\left|
{\omega _-} \right|^2}}} \right] ,\EQN WeylHKX;c\cr}$$
where $\Re$ and $\Im$ stand respectively for the real part and the imaginary part of 
the  argument and $(\quad)^*$ represents the complex conjugate of 
$(\quad)$. $k$ is a constant of integration which is fixed by the boundary conditions at 
infinity.
The  two yet unspecified function $\beta$ and $\varpi$ satisfy:
$$\EQNalign{&\nabla \beta (a_k)={1 \over {r_k}}\left(
{\left[ {z-a_k} \right]\nabla +R\,{\bf e}_\phi \times \nabla } \right)\,\nu_0 ,\EQN 
beta;a \cr
  &\nabla \varpi =R\,{\bf e}_\phi \times \nabla \nu_0, \EQN beta;b\cr}$$
 where $ \nabla =( { \partial / \partial R },{ \partial  / \partial z } )$
  and ${\bf e}_\phi$ is the unit vector in the $\phi$ direction.
 The linearity of these equations suggests again that solutions should be sought in terms of 
 the line  density, $W(b)$, characterising $\nu_0$, namely 
$$ \EQNalign{
 \beta(a_k) &= \int { W(b) \over a_k }  
\left[{ R^2 +  (z-b-a_k)^2 \over R^2 + (z-b)^2 }\right]^{\half}  db, \EQN  betap;a \cr 
 \varpi &=\int    {W(b) (b-z) \over \sqrt{ R^2 + (z-b)^2}} \, db. \quad \quad \quad 
\EQN betap;b  \cr }$$
It is therefore a matter of algebraic substitution to apply the above
procedure and construct all non-linear stationary vacuum fields of the
Papapetrou form from static Weyl fields.  Following the prescription
described in section~2, the corresponding disk in real rotation is
then constructed.  The requirement for the physical source to be a
disk is in effect a less stringent condition on the regularity of the
vacuum metric in the neighbourhood of its singularity because only the
half space which does not contain these singularities is physically
meaningful.  A suitable Ehlers transformation on $D_-/D+$ ensures
asymptotic flatness at large distance from the source.

To illustrate this prescription consider the vacuum 
field given by Yamazaki (1981)\cite{Yamazaki}.
This field arises from 2 HKX rank zero transformation on the
 Zipoy-Voorhess\cite{Voorhess} static metric
given by  
 $$\EQNalign{\nu_0 &= \half \delta \log{{(x-1)\over (x+1)}},\EQN yama;a \cr
 \zeta_0 &= \half \delta^2 \log{{(x-1) \over (x^2-y^2)}}, \EQN yama;b\cr }$$
in prolate spheroidal co-ordinates. 
It corresponds to a uniform line density between $\pm \delta$. The functions
$\beta$ and $\varpi$ follow from \Eqs{betap}.
Choosing poles $a_k, \, \, k=1,2$ at the end of the rod $\pm \delta$,
leads to
$$\EQNalign{ \beta(a_\pm) &= \half \delta \log{{x^2-1 \over \left[ x \mp y\right]^2}}, 
\EQN betaY;a \cr
	\varpi &= 2 \delta y.\EQN betaY;b \cr }$$
Equations \Ep{WeylHKX} together with \Ep{yama} and \Ep{beta} characterize completely the three 
Weyl-Papapetrou metric functions $\nu,\zeta, \, {\rm and} \,
\omega$;  all physical properties of the corresponding disks follow.
For instance, \Fig{Yamazaki} gives  the zero radial pressure velocity curve of the Yamazaki
disk spanning from the Schwarzschild ($\delta =1/2$) metric when $\alpha_1=\alpha_2$.

More generally, the  extension of this work to the construction of disks with planar
 {\sl isotropic }
pressures should be possible by requiring that \Eq{FinalV;c} is identically 
 equal to \Eq{FinalV;d}  given \Eq{WeylHKX}.
\figure{Yamazaki}
\Caption
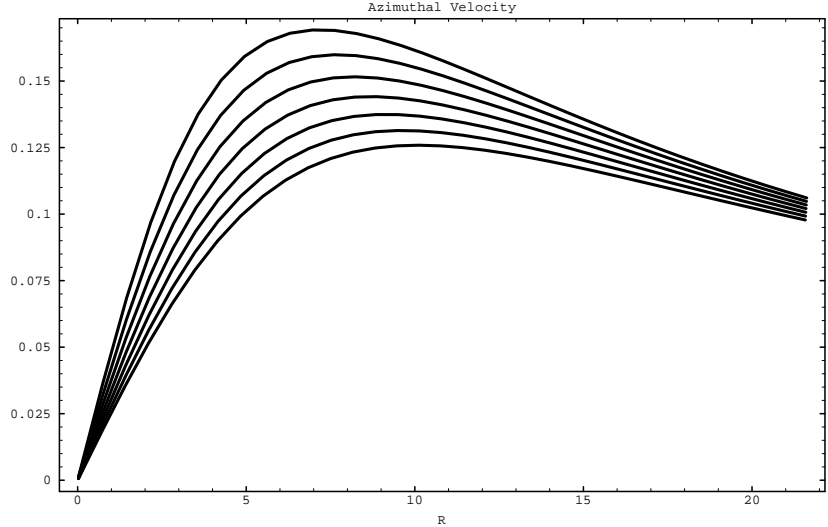
 the azimuthal velocity for the  $a/\kappa_Y$ zero radial pressure 
Yamazaki disk as a function of circumferential radius 
$R_c$,
for a class of solutions with decreasing  potential compactness
$b/\kappa_Y = 3.6,3.8, \cdots 5$. $\kappa_Y$ is the natural unit of 
distance in these disks as defined by Yamazaki's  Eq. (Y-2)\cite{Yamazaki}.  
\endCaption
\endfigure 

\section{ Conclusion}
  
The general counter rotating disk with partial pressure support has
been presented.  The suggested method for implementing radial pressure
support also applies to the construction of stationary axisymmetric
disk with rotation but requires prior knowledge of the corresponding
vacuum solution.  In this manner, a disk-like source for the Kerr
field has been constructed orbit by orbit.  The corresponding
distribution reduces to a Keplerian flow in the outer part of the disk
and presents strong relativistic features in the inner regions such as
azimuthal velocities close to that of light, large central redshift
and ergoregions. The disk itself is likely to be stable against ring
formation and the ratio of its binding energy to rest mass can be as
large as 1:10.  The broad lines of how to construct all rotating disks
arising from HKX-transforming the corresponding counter rotating model
into a fully self-consistent model with proper rotation and partial
pressure support has been sketched.  It should be a simple matter to
implement this method with the additional requirement that the
pressure remains isotropic with a sensible polytropic index. One could
then analyse the fate of a sequence of gaseous disks of increasing
compactness and relate it to the formation and evolution of quasars at
high redshift.  Alternatively, the inversion method described in
section~4.1 yields a consistent description of the detailed dynamics for
all disks in terms of stellar dynamics.

\nosechead{ References }

\ListReferences
\vfill
\eject

\appendix{A}{ Equation of state for a relativistic  2D adiabatic flow }
The equation of state of a planar adiabatic flow 
is constructed here by relating the perfect fluid stress energy tensor  of the flow to 
 the most probable distribution function  which  maximises 
the Boltzmann entropy.\Note{ Synge\cite{Synge} gives an extensive derivation 
 of the corresponding  equation of state for a 3D flow.}
These properties are local characteristics of the flow. It is therefore assumed that
all tensorial quantities introduced in this section are expressed in the local Vierbein
frame.

Consider a infinitesimal volume element $d \vartheta$  defined in the neighbourhood of 
a given event, and assume that the particles entering 
this volume element are subject to molecular chaos. 
The distribution function $F$ of particles is defined so that 
$F({\bf R}, {\bf P}) \, \, d\vartheta d^2 P$ gives the number of particles in the 
volume $\vartheta$ centered on $\bf R$ with 2-momentum pointing 
to $\bf P$ within $d^2P$.

The most probable distribution function $F^\star$ for these particles
is then given by that which maximises Boltzmann entropy $S$ subject to the  constraints 
imposed by the conservation of the total energy momentum and by the conservation of the number of 
particles within that volume.
This entropy reads in terms of the distribution $F$
$$ S = - d \vartheta \int F \log F  \, d^2 P \,.  \EQN entropy1 $$
The stress energy tensor corresponds to the  instantaneous 
flux density of energy momentum through 
the elementary volume  $d \vartheta$ (here a surface)
$$ T^{\alpha b}=   \int  { F^\star P^\alpha P^\beta} \,{ d^2 P  \over| {\cal E}| }\, , \EQN tab $$
where the $1/|{\cal E}|$ factor accounts for the integration over energy-momentum space 
to be restricted to  the pseudo-sphere $ P^\alpha P_\alpha = m^2$.
Indeed the detailed energy-momentum conservation requires the integration to be carried along
the volume element
$$ \int \delta \left( P^\alpha P_\alpha - m^2\right) \, d^3 P =
\int \delta \left( {P^0}^2 - {\cal E}^2\right) \, d^2 P  dP^0 =  
 { d^2 P \over{|\cal E}|}  \, ,
 \EQN ugly $$  where ${\cal E} = \sqrt{ m^2 +  {\bf P}^2}$. It is assumed here that momentum 
space is locally flat.
Similarly the numerical flux vector reads
$$ \phi^\alpha_\star = \int F^\star P^\alpha  \,{ d^2 P \over |{\cal E}|}\, . \EQN  flux $$ 
The  conservation of energy-momentum of the volume element $d \vartheta$ implies 
that the flux of energy momentum across
that volume should be conserved; this flux reads
$$  T^{\alpha \beta} \, n_\beta\, d \vartheta =  d \vartheta 
\int   F^\star P^\alpha \, d^2 P  ={\rm Const.} \, , \EQN c1 $$
where $ n_\beta$ is the unit  time axis vector.
Keeping the population number constant provides the last  constraint on 
the possible variations of $S$
$$ \phi_\star^\alpha n_\alpha d \vartheta  = d \vartheta \int  F^\star d^2P ={\rm Const.} \, .\EQN c2 $$

Varying \Eq{ entropy1} subject to the constraints \Ep{c1} \Ep{c2}, and 
putting $\delta S$ to zero   leads to
$$ (\log F^\star + 1) \delta  F^\star = a \delta  F^\star + \lambda_\alpha P^\alpha 
 \delta  F^\star \, , \EQN $$
where $a$ and $\lambda_\alpha$ are Lagrangian multipliers corresponding resp. to 
\Ep{c1} \Ep{c2}. These multipliers are independent of $P^a$ but in general will be 
a function of position.
The distribution which extremises $S$ therefore reads
$$ F^\star = C  \exp( \lambda_\alpha P^\alpha ) \, . \EQN $$ 
The 4 constants $C$ and $\lambda_a$ are  in principle fixed by 
\Eqs{c1} and \Ep{c2}.
The requirement for $F^\star$ to be Lorentz invariant -- that is independent of 
the choice of  normal $n_\alpha$ -- and 
self consistent, 
 is met instead by  demanding that $F^\star$ obeys \Eq{tab} and \Ep{flux} , namely
$$ \EQNalign{& C \, \int P^\alpha  \exp( \lambda_\mu P^\mu ) {d^2P \over  |{\cal E}|} =  \phi_\star^\alpha \, ,  \EQN c3 \cr
& C \, \int P^\alpha P^\beta  \exp( \lambda_\mu P^\mu ) {d^2P \over  |{\cal E}|} =  T^{\alpha \beta}\, ,  \EQN  c4 \cr} $$
where $\phi_\star^\alpha $ and $T^{\alpha \beta}$ satisfy in turn the covariant constraints 
$$\EQNalign{ 
{\partial T^{\alpha \beta} \over {\partial x^\beta} } &= 0 \quad {\rm  and} \quad  \EQN c5;a \cr 
 { \partial \phi_\star^\alpha \over \partial x^\alpha}&=0 \, .  \EQN c5;b  \cr }$$
Equations \Ep{c3}-\Ep{c5}  provide a set of  thirteen equations to constraint the thirteen functions
$C, \lambda_\alpha, \phi_\star^\alpha,\,{\rm and} \,\, T^{ \alpha \beta} $.
Equations \Ep{c3} and \Ep{c4} may be rearranged as
$$ \phi_\star^\alpha = C { \partial \Phi  \over \partial \lambda^\alpha }\, , \quad
T^{\alpha \beta}=  C { \partial^2 \Phi \over \partial \lambda^\alpha  \partial \lambda^\beta}\, , \EQN  $$
where the auxiliary function $\Phi$ is defined as $\Phi = 
\int  \exp( \lambda_\alpha P^\alpha )
 \, 
d^2 P /|{\cal E}| $;
$\Phi$ is best evaluated using pseudo polar coordinates  corresponding to
 the symmetry imposed by the energy momentum 
conservation $ P^\alpha P_\alpha - m^2=0$; these are
$$ \eqalign{&  P^1 = m \sinh \chi  \cos \phi\, , \cr 
	&  P^2 = m \sinh \chi\sin \phi \, ,\cr 
&  P^0 =  i m \cosh \chi \, , \cr 
 } \EQN $$	   
where the coordinate $\chi$ is chosen  so that the normal $n_\alpha$  lies along $\chi =0$. 	
In terms of these variables, $d^2P /|{\cal E} | $ then reads
$$ d^2P/ |{\cal E }| = d P^1 dP^2/m  \cosh \chi   =  m \sinh \chi  d \chi d \phi \, . \EQN $$
Moving to a temporary frame in which $\lambda^0$ is the time axis ( $\lambda^0 = i\lambda =
i (-\lambda_\alpha \lambda^\alpha)^{1/2},$ $ \,\,  \lambda^k = 0 \, \, k=1,2$), $\Phi$ becomes
$$ \Phi = 2 \pi m \int^{\infty}_0 \exp(- m \lambda \cosh \chi) \sinh \chi d \chi=
{ 2 \pi \over  \lambda}  \exp(-  \lambda) \, . \EQN eqPhi$$
From \Eq{c3}-\Ep{c4} and \Ep{eqPhi}, it follows that 
$$\EQNalign{ &
 T^{\alpha \beta} =  { 2 \pi C \over  \lambda^3 }  
\exp(-  \lambda) \left[ \left\{  \lambda^2 + 3 \lambda + 3 \right\} {\lambda^\alpha \over \lambda}
 {\lambda^\beta \over \lambda}
+ \left\{  \lambda +1 \right\}  
\delta^{ \alpha \beta}  \right] \, ,  \EQN t1 \cr
& \phi_\star^{\alpha } = 
 { 2 \pi  C \over  \lambda^3 }  \exp(-  \lambda) \left\{  \lambda +1 \right\}  \lambda^\alpha
 \, .  \EQN  phi1 \cr  } $$
On the other hand, the stress energy tensor of a perfect fluid is 
$$ T^{\alpha \beta} = ( \varepsilon_0 + p_0) {U^\alpha}  {U^\beta}
+ p_0 \, \delta^{ \alpha \beta} \, ,  \EQN  t2$$
while the numerical surface density of particles measured in 
the rest frame of the fluid $\Sigma$ is related to $  \phi_\star^{\alpha } $ 
via 
$$ \Sigma = -  \phi_\star^{\alpha } U_\alpha \, . \EQN  sigma1 $$
\Eq{t1} has clearly the form of \Eq{t2} when identifying
$$ \EQNalign{U^\alpha  &= \lambda^\alpha /\lambda \, ,  \EQN ident;a \cr 
      \varepsilon_0 + p_0 &=    { 2 \pi C \over  \lambda^3 }  \exp(-  \lambda) \left\{  \lambda^2 + 3 \lambda +3 \right\} \, ,  \EQN ident;b\cr
 		    p_0 &=    { 2 \pi C \over  \lambda^3 }  \exp(-  \lambda)  \left\{  \lambda+1  \right\}  	
\, .  \EQN ident;c  \cr }$$
Eliminating $\lambda$,$C$  between \Eqs{phi1}--\Ep{ident;c}
yields the sought after equation of 
state
  $$\varepsilon_0  =  2 \, p_0 + {\Sigma \over 1+ p_0/\Sigma } \, .  \EQN eos$$
where  $p_0 = \Sigma / \lambda $. This is the familiar gas law given 
that $1/\lambda$ is the absolute temperature.
The equation of state \Eq{eos} corresponds by construction to an isentropic
flow. Indeed, using the conservation  equations \Eq{c5;a} dotted with $U^\alpha$,
 and \Eq{c5;b} together with
$ U^\alpha U_\alpha =-1 $ yields  after some algebra 
to the identity
$$\lambda { d   \over d s}\left[ (\lambda^2 + 3 \lambda+3)\over \lambda (\lambda+1) \right] +  {1 \over \lambda} =
 -{ \partial U^\alpha \over \partial x^\alpha} =   {1\over \Sigma } 
 { d  \Sigma \over d s}\, , \EQN  dG$$
where ${ d  /d s} = ({ d x^\alpha / d x^\alpha}) ({ \partial  / \partial x^\alpha} )$
is the covariant derivative following the stream lines.
Writing the {\it r.h.s} of \Eq{dG } as an exact covariant derivative leads to the first 
integral
$$  { d   \over d s}\left[ {1\over {1 + \lambda }} - 2\,\log (\lambda ) + \log (1 + \lambda ) 
- \log(\Sigma) \right] =0 \, . \EQN $$
Given that  $p_0 = \Sigma / \lambda $, it follows  
$$ (1+ { p_0 \over\Sigma} ){p_0 \over\Sigma} \,  \exp\left({ 1 \over  1+ \Sigma/p_0  }\right)
 {1 \over\Sigma}
= {\rm Const.} \EQN $$
which in the low temperature limit gives $\Sigma^{-2} p_0= {\rm Const.}$ This corresponds to
 the correct adiabatic index $\gamma = (2 +2)/2$.

\appendix{B}{ Distribution functions  for counter-rotating relativistic  disks }
For counter rotating disks in the absence of gravomagnetic forces, the 
inversion for distribution functions given in section~4.1 
can be carried in close analogy to the Newtonian method given by 
Pichon \& Lynden-Bell (1995)\cite{PDLB}, and is sketched here. 
The relativistic flow is characterised by its 
stress energy tensor, $T^{(\alpha \beta )}$, which, in Vierbein 
components,   reads
$$T^{(\alpha \beta )}=\int {\int {{{f_{\star}(\epsilon,h)\,P^{(\alpha )\,}P^{(\beta )}\,} \over
 {P^{(t)}}}}}\,dP^{(R)}dP^{(\phi )} \, . \EQN Tf$$
From the geodesic equation it follows that
$$P^\mu P_\mu =-1\quad \Rightarrow \quad P^{(R)}={1 \over
 {\sqrt {\gamma_{tt}}}}\left[ {\epsilon ^2-h^2 \, \gamma_{tt}/
\gamma_{\phi \phi }-g_{tt}} \right]^{1/2} \, , \EQN geo $$
where the line element in the disk is taken to be
$$ds^2 =-\gamma_{t t}\, dt^2 + \gamma_{R R} \,dR^2 + \gamma_{\phi \phi }\, d\phi^2 \, . \EQN metricdisk$$
Differentiating \Eq{geo} implies  $${{dP^{(R)}} \over {P^{(t)}}}=
{{d\epsilon } \over {\left[ {\epsilon ^2-h^2 \,\gamma_{tt}/\gamma_{\phi \phi }
-\gamma_{tt}} \right]^{1/2}}}\, , \quad {\rm while}\quad dP^{(\phi )}={{dh} \over
 {\sqrt {\gamma_{\phi \phi }}}}\, , \EQN $$
where the differentiation is done at constant $R$ and $h$ , and $R$ and $\epsilon$ respectively.
Combinations of \Eqs{Tf} lead to
$$\EQNdoublealign{&\Lambda = {\gamma_{\phi \phi }}^{1/2}\,\left( {T^{(tt)}-T^{(\phi \phi )}-
T^{(RR)}} \right)&=\int {\int {{{f_{\star}(\epsilon,h)\,dh \, d\epsilon \,} 
\over {\sqrt{ {\epsilon ^2-h^2\gamma_{tt}/\gamma_{\phi \phi }-
\gamma_{tt}} }}}}}\, , \EQN tol;a\cr
  &\Delta = {\gamma_{\phi \phi }}^{3/2} \,T^{(\phi \phi )}&=\int {\int
{{{h^{2\,}f_{\star}(\epsilon,h)\,dh d\epsilon \,} \over {\sqrt{
{\epsilon^2 -h^2\gamma_{tt}/\gamma_{\phi \phi }-\gamma_{tt}}
}}}}}\, . \EQN tol;b\cr}$$ Note that \Eq{tol;a} corresponds to Tolman's
formula\cite{Tolman} which yields be integration the total gravitating mass of the
disk.  Introducing
$${\cal R}^2=\gamma_{\phi \phi }/\gamma_{tt}\,
 , \quad e={1 \over 2}\left( {\epsilon ^2-1} \right)\,,
\quad \Psi ={1 \over 2}\left( {1-\gamma_{tt}} \right)\, \quad {\rm and } \quad 
\hat f_{\star}(e,h)={{f(\epsilon ,h)} \over \epsilon }\, ,  \EQN $$ gives, for \Eqs{tol}:
$$\Lambda =\int {\int {{{\hat f_{\star}(e,h)\,dhde\,} \over {\left[ {2(e+\psi )-h^2/{\cal R}^2} \right]^{1/2}}}}}
\,\quad {\rm and } \quad \Delta =\int {\int {{{h^{2\,}\hat f_{\star}(e,h)\,dhde\,}
 \over {\left[ {2(e+\psi )-h^2/{\cal R}^2} \right]^{1/2}}}}}\,. \EQN tol2$$
\Eq{tol2} is formally identical to Eq.(2.9) of Pichon \& Lynden-Bell\cite{PDLB}, replacing $(\Lambda,\Delta,\hat f_{\star},{\cal R},e,\Psi)$ by 
$(R\, \Sigma, R^{3} \,p_\phi,$ $f,R,\epsilon,\psi)$. 
In fact $(\Lambda,\Delta,\hat f_{\star},{\cal R},e,\Psi)$ tends to 
$(R\, \Sigma, R^{3} \, p_\phi, $ $ f,R,\epsilon,\psi)$ in the classical regime.
Introducing the intermediate functions $\hat F(h,{\cal R})$, which is chosen 
so that its  moments satisfy \Eqs{tol}, and following the classical
 prescription of  Pichon \&  Lynden-Bell\cite{PDLB}
yields
	$$\hat f({e},h) ={1 \over  \pi} \int \limits_{{\cal R}_e}^{{\cal R}_p} 
 {\left({\partial { \hat F } / \partial {\cal R}}\right)_h \,d  {\cal R} \over \sqrt{ h^2/{\cal R}^2 - 2\Psi-2{e} }}-
{1 \over  \pi} \int \limits_{{\cal R}_a}^{\infty} 
{  \left({\partial {\hat  F } / \partial {\cal R}}\right)_h\,d  {\cal R} \over \sqrt{ h^2/{\cal R}^2 - 2\Psi-
2{e} }} \, ,  \EQN invertAbelRG $$
where ${\cal R}_p(h,e)$, and ${\cal R}_a(h,e)$ are respectively the
apogee and perigee of the star with invariants $(h,e)$, and ${\cal
R}_e(h)$ is the inner radius of a star on a ``parabolic'' (zero $e$ energy) orbit with
momentum $h$.
\Eq{invertAbelRG } provides  direct and systematic means to construct families of 
distribution function  for counter-rotating disks characterised by their stress-energy tensor in the plane.
Note that in contrast the dragging of inertial frames requires to fix three
moments of the velocity distribution to account for the given energy
density, pressure law and rotation law. 

\appendix{C}{ Distribution functions for rotating  disks: 
Lorentzian in momentum} 

A possible choice for $\bar F$ is a Lorenzian parametrised in  width, $p_2$,
mean, $p_1$ and amplitude, $p_0$ such that
$$ {\bar F}(v_\phi,R) d v_\phi = {p_0\, p_2 \,\pi^{-1} \, d p \over p_2^2+ (p-p_1)^2 } \, ,
\quad {\rm where } \quad p= {v_\phi \over \sqrt{1-v_\phi^2} } \quad
{\rm and } \quad  v_\phi = e^{\nu} (\chi - \omega)/R_c \, . \EQN fvphiLorentz $$
The explicit function of $v_\phi$ is found using $ d p = d
v_\phi/(1-v^2_\phi)^{3/2}$. 
In this instance, \Eq{T12} becomes
$$ \EQNalign{ 
T^{( t t )}\, e^{4 \nu}   =&
{p_0 \, p_2 \over \pi} \int _{-\infty }^{\infty }{dp \over
\,\left( {{p_2}^2} + {{\left( p - {p_1} \right) }^2}
\right) } = p_0 \, ,  \EQN tlorenz;a \cr 
T^{( t \phi )}\, e^{4 \nu} =&
{p_0 \, p_2 \over \pi} \int _{-\infty }^{\infty }{p\,dp\over 
       {{\sqrt{1 + {p^2}}}\,\left( {{p_2}^2} + {{\left( p - {p_1} \right) }^2}
\right) }} = {\cal G}(p_0,p_1,p_2)\, ,  \EQN tlorenz;b \cr
T^{( \phi \phi )}\, e^{4 \nu} =&
{p_0 \, p_2 \over \pi} \int _{-\infty }^{\infty }{p^2\,dp\over 
       {{(1 + {p^2})^2}\,\left( {{p_2}^2} + {{\left( p - {p_1} \right) }^2}
\right) }} =
{ p_0 {\left( {p_2} + {{p_2}^2} + {{p_1}^2} \right) }\over {
(1+{p_2})^2 + {{p_1}^2}}} \, \EQN tlorenz;c \cr } $$ where ${\cal
G}(p_0,p_1,p_2)$ is given by \Eq{gap} bellow.  The inversion of
\Eqs{tlorenz} for $(p_0,p_1,p_2)$, together with \Eqs{fvphiLorentz},
\Ep{finaldf} and \Eq{Pv;c} yields the distribution function of a
Lorenzian flow compatible with the energy distribution imposed by the
metric $(\nu,\zeta,\omega)$ and the cut $z= f(R)$.  This inversion is
always achievable provided the energy conditions \Eq{constraint} are
satisfied.

 More generally, families of distributions functions corresponding to
azimuthal momentum distribution such as 
 $2 p_2^3 \,  p_0
\, \pi^{-1} /( p_2^2+ (p-p_1)^2 )^2 , 8 p_2^5 \, p_0\,(3 \pi)^{-1} /( p_2^2+(p-p_1)^2 )^3
\cdots$ are readily derived in a similar manner while differentiating
the {\it r.h.s.} of \Eq{tlorenz}
with respect to $p_2^2$. These higher order distributions 
have all vanishing number of stars with azimuthal velocity close to
the velocity of light.
This parametrisation is illustrated for the Kerr disk on 
\Fig{kerrDF}.

\subsection{The quadrature for $  {\cal G}(p_0,p_1,p_2)$ } 
Changing variables to $u$ such that $p=  (u-1/u)/2$ 
\Eq{tlorenz;b } becomes
$$   {\cal G}(p_0,p_1,p_2)=
{2 p_0 \, p_2 \over \pi} \int _{0}^{\infty }{(u^2 -1) \,du\over 
       4 {{p_2}^2}\, u^2  + {{\left( u^2 -1 -2  {p_1}\, u \right) }^2}
 } \EQN d2
$$
This integral is then carried and yields
$$ 
\displaylines{   {\cal G}(p_0,p_1,p_2) = 
{{p_0 \left( p_2 + i\,{p_1} \right) \,\log (-2\,i\,{p_2} + 2\,{p_1} -
2\,{\sqrt{1 + {{\left( -i\,{p_2} + {p_1} \right) }^2}}})}\over
{2\,{\sqrt{1 + {{\left( -i\,{p_2} + {p_1} \right) }^2}}}\,\pi }}+ \cr
{{{p_0}\,\left( i\,{p_2} - {p_1} \right) \,\log (-2\,i\,{p_2}
+ 2\,{p_1} + 2\,{\sqrt{1 + {{\left( -i\,{p_2} + {p_1} \right)
}^2}}})}\over  2 {{\sqrt{1 + {{\left( -i\,{p_2} + {p_1} \right)
}^2}}}\,\pi }}+ 
 {p_0{\left( {p_2} - i\,{p_1} \right) \,\log
(2\,i\,{p_2} + 2\,{p_1} - 2\,{\sqrt{1 + {{\left( i\,{p_2} + {p_1}
\right) }^2}}})}\over {2\,{\sqrt{1 + {{\left( i\,{p_2} + {p_1} \right)
}^2}}}\,\pi }}+ \cr {{{p_0}\,\left( i\,{p_1} - {p_2} \right)
\,\log (2\,i\,{p_2} + 2\,{p_1} + 2\,{\sqrt{1 + {{\left( i\,{p_2} +
{p_1} \right) }^2}}})}\over 2 {{\sqrt{1 + {{\left( i\,{p_2} + {p_1}
\right) }^2}}}\,\pi }}  \cr }
  $$ 
$$\EQN gap $$


\appendix{D}{ The relativistic epicyclic frequency}
 The equation for the radial oscillations 
follows from the 
Lagrangian $${\cal L} =  - \sqrt{ 
e^{ 2 \nu }\, \left( {\dot t} -\omega {\dot \phi } \right)^2-
 e^{ 2 \zeta-2 \nu } \, (1+f'^2)\, {\dot R}^2 -  \, e^{- 2 \nu} \,  {\dot \phi }^2 R^2 
}, \EQN lagrange $$ 
where
$\mathop {(\quad)}\limits^\cdot $
 stands here 
for derivation with respect to the proper time
$\tau$ for the star describing its orbit. 
In \Eq{ lagrange}, $\phi$ and $t$ are ignorable which leads to the invariants:
$$
{\partial  {\cal L} \over \partial {\dot \phi } } = h     \quad \quad
{\partial  {\cal L} \over \partial {\dot t} } = -\epsilon. \EQN invariant $$
The integral of the motion for the radial motion follows from \Eqs{invariant} and
  $ U^\mu U_\mu =-1$, namely:
$$  e^{2 \zeta- 2 \nu } \, (1+f'^2) {\dot R}^2 + {e^{ 2 \nu } \over R^2} 
\left(   
h-\epsilon \, \omega
\right)^2 -\epsilon^2 e^{ - 2 \nu} =-1$$
This equation, together with its total derivative with respect to proper time provides
 the radial 
equation of motion, having solved for the  angular momentum $h$ and  the energy $\epsilon$ of
circular orbits as a function of radius when equating both $\dot R$ and $ \ddot R$
to zero.
the relativistic generalisation of the classical 
epicyclic frequency is defined here to be the frequency of  the oscillator  calling back linear
 departure from circular orbits. Hence the equation 
for  radial perturbation reads
$$ \delta {\ddot  R}+ \kappa^2 {\delta R}=0, \EQN radialEQ$$
which gives for $\kappa^2$ :
$$\eqalign{
 \kappa^2 =& e^{-2 Z} \left( Z''-2Z'^2\right) -\epsilon^2 e^{-2 L} 
\left( L''-2L'^2\right)-
            \cr &
e^{2 K} \left[ 
\eqalign{ &
(h-\epsilon \omega)^2(K''+2 K'^2) - \epsilon (h- \epsilon \omega)
 (4 K' \omega' + \omega'') +
 \epsilon^2 \omega'^2 
\cr }\right],
\cr } \EQN kappa $$
where $(\quad)'$ stands in this Appendix for
 $d/dR \equiv  \partial/\partial R + f'\,  \partial/\partial z$
and $L,\,K,\,{\rm and}\, Z$ are given in terms of the potential $\zeta$ and $\nu$ by:
$$ \EQNalign{&
L= \zeta + \log{\sqrt{1+f'^2} },
                    \EQN LNZ;a \cr &
Z= \zeta + \log{\sqrt{1+f'^2} } -\nu,
                   \EQN LNZ;b \cr &
K= 2\nu -\zeta - \log{\sqrt{1+f'^2} } -\log{R}.
                     \EQN LNZ;c \cr &
            } $$
Here $\epsilon$ and $h$ are known functions of $R$ given by the roots of:
$$ \EQNdoublealign{
{\dot R}&= 0 \quad \Rightarrow \quad e^{ -2 L} \epsilon^2 &= (h-\epsilon\, 
\omega)^2 e^{2 N} + e^{-2 Z},
       \EQN Rdot2;a \cr
{\ddot R}&=0 \quad \Rightarrow \quad e^{-2 Z} Z' &= e^{ -2 L} L' \epsilon^2 + 
\EQN Rdot2;b \cr &
& (h-\epsilon \omega) e^{2 N} \left[ (h-\epsilon\, \omega) N' -\epsilon \,\omega' 
\right].
   \cr }$$

\appendix{E}{Correction to Lemos' Self-Similar Disk}
 The result given in section 6.1 agrees with that of Lemos, having made 
the following  agreed corrections to his solution:
\item {$\bullet$} Eq.$(L \, 3.14)$  should read: $ (k+n) \cot\half (k+n)\pi=-4\pi p_{* 
rr}$
with a minus sign, because the integration  of $\theta$ is carried downwards. This 
gives
$n+k \geq 1$ for the disks with positive pressures.
\item {$\bullet$}  Lemos gives the stress energy tensor integrated through the plane 
using
coordinate increment  rather than proper length. Our result gives the latter which has 
an 
extra factor exp($\zeta-\nu$) and gives the physical energy tensor per unit area.
\item {$\bullet$} The exponent of $ \cos[ (n+k) \pi/2]$ is $2n(k-n)/(n+k)^2$ rather 
than 
$4n(1-2n) /(k+n)^2$.

\bye